\documentclass[a4paper,12pt]{iopart}
\makeatletter
\providecommand*\@nameundef[1]{\expandafter\let\csname #1\endcsname\@undefined}
\@nameundef{equation*}
\@nameundef{endequation*}
\makeatother

\usepackage{amsmath}
\usepackage{amssymb}
\usepackage{amsfonts}
\usepackage{amsthm}
\usepackage{mathtools}
\usepackage{lmodern}
\usepackage[T1]{fontenc}

\newtheorem{theorem}{Theorem}
\newtheorem{lemma}[theorem]{Lemma}
\newtheorem{proposition}[theorem]{Proposition}
\theoremstyle{definition}
\newtheorem{definition}{Definition}
\newcommand*{\defof}[1]{\emph{#1}}
\newcommand*{\Lmaref}[1]{Lemma~\ref{#1}}
\newcommand*{\Propref}[1]{Proposition~\ref{#1}}
\newcommand*{\Defref}[1]{Definition~\ref{#1}}

\def\figscale{1.0}
\usepackage[%
    backend = biber, %
    natbib = true, %
    maxnames = 9, %
    style = alphabetic, %
    citestyle=alphabetic, %
    sorting = nyt, %
    giveninits=true, %
    eprint=true, %
    doi=true
]{biblatex}
\newcommand*{\conj}[1]{\overline{#1}}
\newcommand*{\ib}{\bar{\imath}}
\newcommand*{\jb}{\bar{\jmath}}
\let\Laplace\triangle
\newcommand*{\pderiv}[3][]{\frac{\partial^{#1}#2}{{\partial{#3}}^{#1}}}
\newcommand*{\innerProd}[3][]{\left\langle #2, #3 \right\rangle_{#1}}
\newcommand*{\PoiBr}[3][]{\left\{ #2, #3 \right\}^{\mathrm{#1}}}
\newcommand*{\after}{\circ}
\newcommand*{\comm}[3][-]{\left[ #2, #3 \right]_{#1}}
\newcommand*{\bra}[1]{\left\langle #1 \right|}
\newcommand*{\ket}[1]{\left| #1 \right\rangle}
\newcommand*{\braket}[2]{\left\langle #1 \middle\vert #2 \right\rangle}

\usepackage{hyperref}
\hypersetup{%
	pdftitle = {Quantization, dequantization, and distinguished states}, %
	pdfauthor = {Eli Hawkins, Christoph Minz, Kasia Rejzner}, %
	pdfkeywords = {Sorkin-Johnston state, symplectic space, K\"ahler space, geometric quantization, dequantization, Berezin-Toeplitz quantization}, %
	pdfstartpage = 1, %
	pdfstartview = {FitH}, %
	breaklinks = false, %
	colorlinks = true, %
	linkcolor = black, %
	citecolor = black, %
	filecolor = black, %
	menucolor = black, %
	urlcolor = black %
}

\begin{document}

\title{Quantization, dequantization, and distinguished states}

\author{Eli Hawkins$^1$, Christoph Minz$^{1,2}$, Kasia Rejzner$^1$}
% \email{eli.hawkins@york.ac.uk}
% \orcid{https://orcid.org/0000-0003-2054-3152}
% \email{christoph.minz@itp.uni-leipzig.de}
% \orcid{https://orcid.org/0000-0002-5429-5997}
% \email{kasia.rejzner@york.ac.uk}
% \orcid{https://orcid.org/0000-0001-7101-5806}

\address{$^1$ Department of Mathematics, University of York, Heslington, YO10 5DD York, United Kingdom}
\address{$^2$ Institute of Theoretical Physics, Leipzig University, 04103 Leipzig, Germany}
\ead{christoph.minz@itp.uni-leipzig.de}

\begin{abstract}
Geometric quantization is a natural way to construct quantum models starting from classical data. In this work, we start from a symplectic vector space with an inner product and --- using techniques of geometric quantization --- construct the quantum algebra and equip it with a distinguished state. We compare our result with the construction due to Sorkin --- which starts from the same input data --- and show that our distinguished state  coincides with the Sorkin-Johnson state. Sorkin's construction was originally applied to the free scalar field over a causal set (locally finite, partially ordered set). Our perspective suggests a natural generalization to less linear examples, such as an interacting field.
\end{abstract}
\noindent{\it Keywords\/}: Sorkin-Johnston state, symplectic space, K\"ahler space, geometric quantization, dequantization, Berezin-Toeplitz quantization

\submitto{\jpa}

% \maketitle

\section{Introduction}
\label{sec:Introduction}
Geometric quantization provides a very elegant and natural way to quantize classical systems using geometrical data. Despite some limitations, it is still a very attractive approach and in this work we apply it in a new and maybe unexpected context, namely in quantum field theory on \textit{causal sets}.
 
A few years ago, Afshordi, Aslanbeigi and Sorkin \cite{2012AfshordiEtAl} published a construction of a distinguished pure, quasi-free state for quantum field theory in curved spacetimes based on earlier works \cite{2009Johnston,2011Sorkin}.
It was later shown that this state fails to be Hadamard \cite{2012FewsterVerch,2013FewsterVerch}, so its singularity structure does not have some desirable features.
Modifications can recover the Hadamard condition but remove the uniqueness of the state \cite{2014BrumFredenhagen,2018Wingham}. 

In causal set theory one replaces the spacetime manifold with a locally finite partially ordered set, referred to as \textit{the causal set}, for which the Hadamard condition is not meaningful and the Sorkin-Johnston state is applicable without modifications. In this work, we want to study the Sorkin-Johnston state on causal sets from the perspective of geometric quantization.

Traditionally, the Sorkin-Johnston state is obtained as the unique state that satisfies certain natural requirements
\cite{2017Sorkin}, referred to as \textit{Sorkin-Johnston axioms}.
Our construction takes a new perspective (without using the axioms) and delivers the same state through the application of geometric quantization for symplectic manifolds with a Riemannian metric. 
Our construction may later be generalized to an interacting field on causal sets or to quantum field theory on curved spacetime, as long as the required geometry conditions are met.
The main results have also been part of the Ph.D.\ thesis~\cite{2021Minz}.

We review deformation quantization and the Weyl algebra in \Sref{sec:DeformationQuantizationAndSJState}. 
Taking the algebraic perspective, we formulate the Sorkin-Johnston state as a quasi-free state on the Weyl algebra for scalar fields on a causal set.  
In \Sref{sec:GeometricQuantization}, we review geometric quantization and the Toeplitz quantization map, as well as the dual map, known as Berezin-Toeplitz dequantization.  
We apply geometric construction in the case of scalar field theory on a causal set in \Sref{sec:VectorSpace.GeometricQuantization} and show that the dequantization map gives rise to the Sorkin-Johnston state in \Sref{sec:SJStateAndStrictDeformationQuantization}. 
Finally, we show how the Berezin-Toeplitz quantization and dequantization maps correspond to strict deformation quantizations.

\section{Deformation quantization of classical field algebras, and states}
\label{sec:DeformationQuantizationAndSJState}
First, we consider scalar field theory on a causal set to define a vector space with Poisson bracket and determine the corresponding symplectic form. 
For the review of deformation quantization, the Weyl relations as well as the Weyl algebra here, it suffice to know the Poisson structure. 
The symplectic form is required for the derivation of the Sorkin-Johnston state and the geometric construction later on.

\subsection{Scalar fields on causal sets}
\label{subsec:ScalarFieldsCausalSets}
In the algebraic formulation of classical scalar fields on a (finite subset of a) causal set $\mathcal{C}$ as considered in \cite{2020DableheathEtAl}, we start with the \emph{off-shell configuration space} (a vector space) $\mathcal{E}$ of real-valued functions over $\mathcal{C}$ that is equipped with an inner product $\left\langle \cdot, \cdot \right\rangle$.
We are interested in those real-valued functions over $\mathcal{C}$ that obey a discretized version of the Klein-Gordon field equations.

Given the Pauli-Jordan operator $E_{\mathrm{off}} \in \operatorname{End}( \mathcal{E} )$ as the difference of the retarded and advanced Green's operators for the field equations, the space of classical observables $\mathfrak{P}_{\mathrm{off}}$ is the space of complex-valued functions over the configuration space $\mathcal{E}$ with a Poisson structure $\pi_{\mathrm{off}} \in \bigwedge^{2} \mathcal{E}$ so that for all $f_1, f_2 \in \mathfrak{P}_{\mathrm{off}}$:
\begin{align}
\label{eq:OffShellPoissonAlgebra.PoissonBracket}
			{\PoiBr{f_1}{f_2}}_{\mathrm{off}}
	= \pi_{\mathrm{off}}( \mathrm{d}{f_1}, \mathrm{d}{f_2} )
	.
\end{align}
Note that this bracket is equivalently expressed as the map $\pi_\mathrm{off}^\sharp : \mathcal{E}^* \to \mathcal{E}$.
In general, $\pi_\mathrm{off}^\sharp$ is degenerate but its image is an even dimensional sub-space $\mathcal{S} = \operatorname{img}( \pi_\mathrm{off}^\sharp ) \subset \mathcal{E}$.
The \emph{on-shell Poisson algebra} $\mathfrak{P}$ is then the quotient of $\mathfrak{P}_{\mathrm{off}}$ by the ideal generated by all observables that vanish on $\mathcal{S}$ \cite{2020DableheathEtAl}.
We write the Poisson bracket on $\mathfrak{P}$ as $\PoiBr{\cdot}{\cdot}$ and note that the corresponding map $\pi^\sharp : \mathcal{S}^* \to \mathcal{S}$ is now non-degenerate.
The inner product on $\mathcal{E}$ restricts to an inner product $\left\langle \cdot, \cdot \right\rangle$ on $\mathcal{S}$ and determines the metric $g^\flat : \mathcal{S} \to \mathcal{S}^{*}$.
The inverse of the Poisson bracket is a symplectic form $\omega$ on $\mathcal{S}$, which we express with the inverse of the (restricted) Pauli-Jordan operator $E \in \operatorname{End}( \mathcal{S} )$,
\begin{align}
\label{eq:VectorSpace.StructureRelation}
			\forall v_1, v_2
	&\in \mathcal{S}:
	&
			\omega( v_1, v_2 )
	&= \innerProd{v_1}{E^{-1} v_2}
	.
\end{align}

This structure on the vector space $\mathcal{S}$ is our starting point for the state construction.
For a globally hyperbolic spacetime manifold, a symplectic vector space is similarly constructed from the configuration space of smooth functions.
In that case, the symplectic vector space is infinite dimensional.
However, our main focus in this work lies on the given structure for a finite dimensional vector space.

The operator $E$ is anti-symmetric and anti-self-adjoint, $E^* = -E$.
As we have constructed a symplectic vector space, the kernel of the Pauli-Jordan operator $E$ (restricted to $\mathcal{S}$) is trivial and we have the polar decomposition
\begin{subequations}
\label{eq:PauliJordan.PolarDecomposition}
\begin{align}
			E
	&= | E | U^*
	, \\
			E^*
	&= U | E |
	.
\end{align}
\end{subequations}
where $U$ is a unitary operator and $| E |$ is the strictly positive operator i.e., invertible
\begin{align}
\label{eq:PauliJordan.Modulus}
			| E |
	&:= \sqrt{E^* E}
	= \sqrt{- E^2}
	.
\end{align}
We insert this decomposition into \eref{eq:VectorSpace.StructureRelation} to find
\begin{align}
\label{eq:VectorSpace.StructureRelationDecomposed}
			\forall v_1, v_2
	&\in \mathcal{S}:
	&
		- \omega( v_1, U v_2 )
	&= \innerProd{v_1}{| E |^{-1} v_2}
	.
\end{align}
Since $| E |^{-1}$ is a positive, self-adjoint operator, the right hand side of \eref{eq:VectorSpace.StructureRelationDecomposed} is also a symmetric, bi-linear form.
We denote this form by $\eta$,
\begin{align}
\label{eq:VectorSpace.SymmetricForm}
			\forall v_1, v_2
	\in \mathcal{S}:
	\qquad
			\eta( v_1, v_2 )
	:= \innerProd{v_1}{| E |^{-1} v_2}
	.
\end{align}
The operator $J = -U$ is a complex structure on $\mathcal{S}$ (such that $J^2 = -\mathbf{1}$) and the relation between $\omega$ and $\eta$ is K\"ahler.
\begin{definition}
	A \defof{K\"ahler vector space} is a quadruple $( \mathcal{S}, \omega, \eta, J )$ of a vector space $\mathcal{S}$ with a complex structure $J$, a symmetric bi-linear form $\eta$, and a symplectic form $\omega$ such that
	\begin{align}
	\label{eq:VectorSpace.Kaehler}
				\forall v_1, v_2
		&\in \mathcal{S}:
		&\qquad
				\omega( v_1, J v_2 )
		&= \eta( v_1, v_2 )
		.
	\end{align}
\end{definition}
Note that, on the one hand, in the presence of a complex structure $J$, the real vector space $\mathcal{S}$ turns into a complex vector space $\mathcal{S}_{J}$ by
\begin{align}
\label{eq:VectorSpace.ComplexStructure}
			\forall v
	\in \mathcal{S}:
			\forall a, b
	&\in \mathbb{R}:
	&
			 ( a + \mathrm{i} b ) v 
	&:= a v + b J v
	.
\end{align}
with $\dim_{\mathbb{C}} \mathcal{S}_{J} = \frac{1}{2} \dim_{\mathbb{R}} \mathcal{S}$.
On the other hand, the complexification of $\mathcal{S}$ yields a complex vector space $\mathcal{S}^{\mathbb{C}}$ such that $\dim_{\mathbb{C}} \mathcal{S}^{\mathbb{C}} = \dim_{\mathbb{R}} \mathcal{S}$.
The complexified vector space has a holomorphic and an anti-holomorphic subspace, $\mathcal{S}^{\mathbb{C}, +}$ and $\mathcal{S}^{\mathbb{C}, -}$, respectively,
\begin{align}
\label{eq:VectorSpace.ComplexSubspaces}
			\mathcal{S}^{\mathbb{C}, \pm}
	&:= \left\{ v \mp \mathrm{i} J v \,\middle\vert\, v \in \mathcal{S} \right\}
	.
\end{align}
For the introduction of the Sorkin-Johnston in \Sref{subsec:SorkinJohnstonStateAlgebraic} state as well as for its geometric construction in \Sref{sec:VectorSpace.GeometricQuantization} and \Sref{sec:SJStateAndStrictDeformationQuantization}, we use the complex vector space $\mathcal{S}_{J}$ (or equivalently the holomorphic subspace $\mathcal{S}^{\mathbb{C}, +}$). 

\subsection{Formal deformation quantization}
\label{subsec:FormalDeformationQuantization}
Formal deformation quantization is a deformation of the classical pointwise product of the Poisson algebra $\mathrm{C}^\infty( \mathcal{M}, \mathbb{C} )$ to a star product (a series expansion in powers of $\hbar$).
The additional properties of the definition are as in  \cite{2010Schlichenmaier}.
\begin{definition}
\label{def:StarProduct}
	Let $( \mathcal{M}, \{ \cdot, \cdot \} )$ be a Poisson manifold.
	A \defof{star product} $\star$ is a product on the space of formal power series $\mathrm{C}^\infty( \mathcal{M}, \mathbb{C} )[[\hbar]]$ such that for functions $f_1, f_2 \in \mathrm{C}^\infty( \mathcal{M}, \mathbb{C} )$:
	\begin{align}
	\label{eq:StarProduct}
				f_1 \star f_2
		&= \sum_{k = 0}^{\infty} B_{k}( f_1, f_2 ) \hbar^k
		,
	\end{align}
	where $B_{k}$ are bi-linear maps fulfilling the conditions, $\forall f_1, f_2, f_3 \in \mathrm{C}^\infty( \mathcal{M}, \mathbb{C} )$:
	\begin{subequations}
	\begin{align}
	\label{eq:StarProduct.Associativity}
				\text{associativity:} &
		& ( f_1 \star f_2 ) \star f_3
		&= f_1 \star ( f_2 \star f_3 )
		,
		\allowdisplaybreaks[3]\\\label{eq:StarProduct.PointwiseProduct}
				\text{pointwise product:} &
		& B_{0}( f_1, f_2 )
		&= f_1 f_2
		.
	\end{align}
	The star product $\star$ is
	\begin{align}
	\label{eq:StarProduct.PoissonBracket}
				\text{\defof{Poisson compatible} if} &
		& B_{1}( f_1, f_2 ) - B_{1}( f_2, f_1 )
		&= \mathrm{i} \PoiBr{f_1}{f_2}
		,
		\allowdisplaybreaks[3]\\\label{eq:StarProduct.Identity}
				\text{\defof{unital} if} &
		& f_1 \star 1
		= 1 \star f_1
		&= f_1
		,
		\allowdisplaybreaks[3]\\\label{eq:StarProduct.Parity}
				\text{\defof{self-adjoint} if} &
		& \conj{f_1 \star f_2}
		&= \conj{f_2} \star \conj{f_1}
		,
	\end{align}
	\end{subequations}
	and it is \defof{differential} if all $B_{k}$ are bi-differential maps.
\end{definition}

As a standard example, consider the Weyl algebra over a real vector space $\mathcal{S}$ with a Poisson structure $\PoiBr{\cdot}{\cdot}$. 
Let $\mathcal{S}^* = \operatorname{Hom}( \mathcal{S}, \mathbb{R} )$ be the dual vector space. 
The Weyl algebra $\mathfrak{W}_{\hbar}$ is generated by a map $W_{\hbar}$ on $\mathcal{S}^*$ that fulfills the Weyl relations (for all $\forall \phi, \phi' \in \mathcal{S}^*$)
\begin{subequations}
\label{eq:WeylRelations}
\begin{align}
\label{eq:WeylRelations.Product}
            W_{\hbar}( \phi ) W_{\hbar}( \phi' )
    &= \exp\left(
        - \frac{\mathrm{i} \hbar}{2}
            \PoiBr{\phi}{\phi'}
        \right)
        W_{\hbar}( \phi + \phi' )
    ,
\allowdisplaybreaks[3]\\
\label{eq:WeylRelations.Involution}
            W_{\hbar}( \phi )^*
    &= W_{\hbar}( -\phi )
    ,
\allowdisplaybreaks[3]\\
\label{eq:WeylRelations.Unit}
            W_{\hbar}( 0 )
    &= \mathbf{1}
    .
\end{align}
\end{subequations}
These relations are realized by a deformation of $\mathrm{C}^\infty( \mathcal{S}, \mathbb{C} )$ with the Moyal product (exponentiated Poisson bracket followed by pointwise multiplication $m$) 
\begin{subequations}
\label{eq:MoyalProduct}
\begin{align}
            \mathrm{e}^{\mathrm{i} \phi} \star_{W} \mathrm{e}^{\mathrm{i} \phi'}
    &= m \after \exp\left( \frac{\mathrm{i} \hbar}{2} \PoiBr{\cdot}{\cdot} \right)
          \left( \mathrm{e}^{\mathrm{i} \phi} \otimes \mathrm{e}^{\mathrm{i} \phi'} \right)
    ,
\\
    &= \exp\left(
        - \frac{\mathrm{i} \hbar}{2}
          \PoiBr{\phi}{\phi'}
        \right)
        \mathrm{e}^{\mathrm{i} ( \phi + \phi' )}
    .
\end{align}
\end{subequations}
There exists a norm on the image of $W_{\hbar}$ that satisfies the C*-property, so that the image can be completed to a C*-algebra $\mathfrak{W}_{\hbar}$ (that is a subset of bounded operators on some Hilbert space), see \cite{2013Moretti}.

In general, the power series \eref{eq:StarProduct} does not converge, hence the deformation quantization is called \emph{formal}. 
Further below, we determine star products that correspond to Toeplitz quantization and Berezin-Toeplitz dequantization, for which we use the \emph{strict} notion of deformation quantization.

\subsection{Strict deformation quantization}
\label{subsec:StrictDeformationQuantization}
Strict deformation quantization involves a family of C*-algebras $\mathfrak{A}_{\hbar}$ parametrized by $\hbar \in I$ for some parameter range $I \subseteq \mathbb{R}$ including the classical limit $\hbar = 0$.
In the following, we will have $I = [ 0, \infty )$ and denote $I_* = I \setminus \{ 0 \}$.
For some constructions of geometric quantization, however, we will see that the quantization parameter has to take values in $I_* = \left\{ p^{-1} \middle\vert p \in \mathbb{N}_* \right\}$, where the classical limit $\hbar \to 0$ is equivalent to $p \to \infty$.

At the value $\hbar = 0$, take a C*-algebra $\mathfrak{A}_{0}$ between functions vanishing at infinity and bounded functions, $\mathrm{C}_0( \mathcal{M}, \mathbb{C} ) \subseteq \mathfrak{A}_{0} \subseteq \mathrm{C}_\mathrm{b}( \mathcal{M}, \mathbb{C} )$ with the supremum norm (for any $f \in \mathfrak{A}_{0}$) 
\begin{align}
\label{eq:ClassicalAlgebra.SupremumNorm}
          \| f \|
	&:= \sup_{x \in \mathcal{M}} | f( x ) |
    .
\end{align}
On a dense Poisson subalgebra $\mathcal{A}_{0} \subseteq \mathfrak{A}_{0}$, we have a Poisson bracket (either explicit or determined by a symplectic form $\omega$) that is closed under complex conjugation. 

Quantization is described as a deformation of this Poisson algebra by a family of linear maps.

\begin{definition}
\label{def:Quantization}
	A \defof{quantization} $Q$ is a family of linear maps from the classical algebra $\mathcal{A}_{0} \subseteq \mathfrak{A}_{0}$ to the C*-algebra of quantum observables $\mathfrak{A}_{\hbar}$ parametrized by $\hbar$,
	\begin{align}
	\label{eq:Quantization}
					Q_{\hbar} : \mathcal{A}_{0}
		&\to \mathfrak{A}_{\hbar}
	\end{align}
	that respects the involution, $Q_{\hbar}( f )^* = Q_{\hbar}( \conj{f} )$, and if there exists a unit $1 \in \mathcal{A}_{0}$, it is also unital, $Q_{\hbar}( 1 ) = \mathbf{1}$.
\end{definition}
It is easily seen that the map $W_{\hbar}$ that we considered in \Sref{subsec:FormalDeformationQuantization} is a quantization map, known as Weyl quantization. 

In the following, we write the commutator of two operators as $\comm{A}{B} := A B - B A$ and we use the little-o notation i.e., a continuous function $f( \hbar )$ is of order $\operatorname{o}( \hbar )$ if
\begin{align}
\label{eq:Quantization.LittleONotation}
		\lim_{\hbar \to 0}
			\frac{1}{\hbar} f( \hbar )
	&= 0
	.
\end{align}

For a (strict) quantization, one may require additional conditions on the quantization maps, see \cite[ch.~II, Def.~1.1.1]{1998Landsman}.
In particular, we would like to have a quantization that is compatible with the Poisson structure, meaning that for all differentiable functions $f_{1}, f_{2} \in \mathcal{A}_{0}$:
\begin{align}
\label{eq:Quantization.DiracCondition}
			\comm{Q_{\hbar}( f_{1} )}{Q_{\hbar}( f_{2} )}
	&= \mathrm{i} \hbar Q_{\hbar}\bigl( \PoiBr{f_{1}}{f_{2}} \bigr)
		+ \operatorname{o}( \hbar )
    .
\end{align}

For some quantizations, there may also exist a family of dual maps, from the quantum C*-algebras to the C*-algebra of classical observables.

\begin{definition}
\label{def:Dequantization}
	A \defof{dequantization} $\varUpsilon$ is a family of linear maps
	\begin{align}
	\label{eq:Dequantization}
				\varUpsilon_{\hbar} : \mathfrak{A}_{\hbar}
		&\to \mathfrak{A}_{0}
		,
	\end{align}
	that respects involution, $\varUpsilon_{\hbar}( A^* ) = \conj{\varUpsilon_{\hbar}( A )}$, and if there exists a unit $\mathbf{1} \in \mathfrak{A}_{\hbar}$, it is also unital, $\varUpsilon_{\hbar}( \mathbf{1} ) = 1$.
\end{definition}

Note that for a quantization $Q$ and a dequantization $\varUpsilon$, in general $ \varUpsilon_{\hbar} \after Q_{\hbar}$ is not the identity map.

We will consider a continuous field for the family of C*-algebras $( \mathfrak{A}_{\hbar} )_{\hbar \in I}$ over the quantization range $I$ as defined in \cite{1964Dixmier,1977Dixmier}.

\begin{definition}
\label{def:CField}
	Let $I$ be a topological space and let $( \mathfrak{A}_{\hbar} )_{\hbar \in I}$ be a family of C*-algebras.
	A \defof{continuous field of C*-algebras} is a triple $\bigl( I, ( \mathfrak{A}_{\hbar} )_{\hbar \in I}, \varGamma \bigr)$ with $\varGamma \subseteq \prod_{\hbar \in I} \mathfrak{A}_{\hbar}$ such that
	\begin{enumerate}
		\item $\varGamma$ is a linear subspace of $\prod_{\hbar \in I} \mathfrak{A}_{\hbar}$, closed under multiplication and involution,
		\item for every $\hbar \in I$ the set $\{ A( \hbar ) \in \mathfrak{A}_{\hbar} \mid A \in \varGamma \}$ is dense in $\mathfrak{A}_{\hbar}$, and
		\item for every element $A \in \varGamma$ the norm function $n_{A} : I \to \mathbb{R}$,
			\begin{align}
			\label{eq:CField.NormFunction}
						n_{A}( \hbar )
				&:= \| A( \hbar ) \|
			\end{align}
			is continuous, $n_{A} \in \mathrm{C}( I, \mathbb{R} )$, as well as
		\item for any $A' \in \prod_{\hbar \in I} \mathfrak{A}_{\hbar}$, we have $A' \in \varGamma$ if the following condition is fulfilled. For all $\hbar \in I$ and for all real constants $\delta > 0$ there exists a neighborhood $N_{\hbar} \subset I$ of $\hbar$ such that
			\begin{align}
			\label{eq:CField.Continuity}
						\exists A
				&\in \varGamma:
						\forall \hbar'
				\in N_{\hbar}:
				&
						\| A'( \hbar' ) - A( \hbar' ) \|
				&\leq \delta
				.
			\end{align}
	\end{enumerate}
	The elements of $\varGamma$ are called (continuous) \defof{sections} of the field.
\end{definition}

Products and the involution of any  $A_1, A_2 \in \prod_{\hbar \in I} \mathfrak{A}_{\hbar}$ are computed pointwise, $A_1 A_2 : \hbar \mapsto A_1( \hbar ) A_2( \hbar )$ and $A_1^* : \hbar \mapsto A_1( \hbar )^*$, respectively.

If $I$ is locally compact, then the subset of sections $\mathfrak{A} \subseteq \varGamma$ that have a continuous norm function vanishing at infinity, $n_{A} \in \mathrm{C}_0( I, \mathbb{R} )$, is a C*-algebra with the supremum norm
\begin{align}
\label{eq:CField.Norm}
			\| A \|
	&:= \sup_{\hbar \in I} \| A( \hbar ) \|
	.
\end{align}
The triple $( I, ( \mathfrak{A}_{\hbar} )_{\hbar \in I}, \mathfrak{A} )$ is also referred to as a \emph{C*-bundle} \cite{1995KirchbergWassermann}.

\begin{figure}
	\centering
	\includegraphics[scale=\figscale]{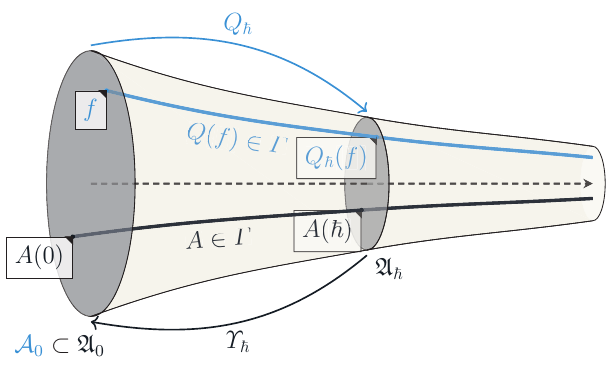}
	\caption{\label{fig:QuantumAlgebraContinuousField} Continuous field of C*-algebras $\bigl( I, ( \mathfrak{A}_{\hbar} )_{\hbar \in I}, \varGamma \bigr)$ with (continuous) sections $Q( f )$ (upper solid line, blue) and $A$ (lower solid line, black), a family of quantization maps $( Q_{\hbar} )_{\hbar \in I}$ and dequantization maps $( \varUpsilon_{\hbar} )_{\hbar \in I}$. At each value $\hbar \in I$ (horizontal axis), there is a quantum algebra $\mathfrak{A}_{\hbar}$ (ellipse, black) with a dense Poisson subalgebra $\mathcal{A}_{\hbar}$ (dotted patterns, blue/gray).}
\end{figure}

As an example for any $f \in \mathcal{A}_{0}$ define
\begin{align}
\label{eq:Quantization.Section}
			Q( f ) : \hbar
	&\mapsto \begin{cases}
				f & \hbar = 0, \\
				Q_{\hbar}( f ) & \hbar \in I_*.
			\end{cases}
\end{align}
Our aim is to determine a quantization such that there exists a continuous field of C*-algebras where these are sections, as sketched in \Fref{fig:QuantumAlgebraContinuousField}.
Furthermore, we want the quantization to admit a star product in the following sense.

\begin{definition}
\label{def:Quantization.StarProduct}
	The \defof{quantization star product $\star_{Q}$} of a quantization --- if it exists --- is the star product \eref{eq:StarProduct} with operators $B_{Q, k} : \mathcal{A}_{0} \times \mathcal{A}_{0} \to \mathcal{A}_{0}$, such that for all $k \in \mathbb{N}$ and for all $f_1, f_2 \in \mathcal{A}_{0}$, the $k$-th order remainder
	\begin{align}
	\label{eq:Quantization.StarProductRemainder}
				R_{Q}^{k}( f_1, f_2, \hbar )
		&:= \frac{1}{\hbar^k}
				\left\|
					Q_{\hbar}( f_1 ) Q_{\hbar}( f_2 )
				- \sum_{j = 0}^{k}
						Q_{\hbar}\bigl( B_{Q, j}( f_1, f_2 ) \bigr)
						\hbar^{j}
				\right\|
	\end{align}
	vanishes in the classical limit,
	\begin{align}
	\label{eq:Quantization.StarProductCondition}
				\lim_{\hbar \to 0}
					R_{Q}^{k}( f_1, f_2, \hbar )
		&= 0
		.
	\end{align}
\end{definition}

Heuristically, the star product of two functions $f_1 \star_{Q} f_2$ of an infinite order quantization $Q$ is an asymptotic expansion of $Q_{\hbar}^{-1}\bigl( Q_{\hbar}( f_1 ) Q_{\hbar}( f_2 ) \bigr)$, though the inverse $Q_{\hbar}^{-1}$ usually does not exist even if the star product exists.

\begin{definition}
\label{def:Quantization.StrictDeformation}
	A quantization $Q$ is an \defof{infinite order strict deformation quantization} if
	\begin{enumerate}
	  \item there exists a continuous field of C*-algebras $\bigl( I, ( \mathfrak{A}_{\hbar} )_{\hbar \in I}, \varGamma \bigr)$ such that
	  	\begin{align}
	  	\label{eq:Quantization.StrictDeformation}
	  				\forall f
	  		&\in \mathcal{A}_{0} \subseteq \mathfrak{A}_{0}:
	  		&
	  				Q( f )
	  		&\in \varGamma,
	  	\end{align}
	  \item the quantization star product $\star_{Q}$ exists, and
	  \item the star product is Poisson compatible.
	\end{enumerate}
\end{definition}

We now consider sections that are well-behaved with respect to quantization and dequantization.

\begin{definition}
\label{def:CField.Expandability}
	A section $A \in \varGamma$ of the continuous field of C*-algebras $\bigl( I, ( \mathfrak{A}_{\hbar} )_{\hbar \in I}, \varGamma \bigr)$ is \defof{$Q$-quantization expandable (quantization expandable or $Q$-expandable for short)} if for any $k \in \mathbb{N}$
	\begin{align}
	\label{eq:CField.QuantizationExpandable}
				\exists f_{0}, \dots, f_{k}
		&\in \mathcal{A}_{0}:
		&
				\lim_{\hbar \to 0}
					\frac{1}{\hbar^{k}}
					\left\| A( \hbar ) - \sum_{j = 0}^{k} Q_{\hbar}( f_{j} ) \hbar^{j} \right\|
		&= 0
	\end{align}
	and it is \defof{$\varUpsilon$-dequantization expandable (dequantization expandable or $\varUpsilon$-ex\-pand\-able for short)} if for any $k \in \mathbb{N}$
	\begin{align}
	\label{eq:CField.DequantizationExpandable}
				\exists f_{0}, \dots, f_{k}
		&\in \mathcal{A}_{0}:
		&
				\lim_{\hbar \to 0}
					\frac{1}{\hbar^{k}}
					\left\|
						\varUpsilon_{\hbar}( A( \hbar ) )
					- \sum_{j = 0}^{k}
							f_{j}
							\hbar^{j}
					\right\|
		&= 0
		.
	\end{align}
	Denote the space of $\varUpsilon$-expandable sections by $\varGamma_{\varUpsilon} \subseteq \varGamma$ and let $\Sigma^{k}_{\varUpsilon}( \cdot, \hbar ) : \varGamma_{\varUpsilon} \to \mathcal{A}_{0}[[ \hbar ]]$ map to the expansion of any section $A \in \varGamma_{\varUpsilon}$ by the functions from \eref{eq:CField.DequantizationExpandable} truncated at order $k$,
	\begin{align}
	\label{eq:CField.DequantizationExpansionMap}
				\Sigma^{k}_{\varUpsilon}( A, \hbar )
		&:= \sum_{j = 0}^{k} f_j \hbar^{j}
		.
	\end{align}
\end{definition}

On the one hand, it is immediately seen that the quantization section $Q( f ) \subset \varGamma$ for any $f \in \mathcal{A}_{0}$ is $Q$-expandable with $f_{0} = f$ and $f_{k} = 0$ for all $k > 0$.
On the other hand, if the space of $\varUpsilon$-expandable sections forms a *-subalgebra of $\varGamma$, then dequantization also admits a star product (which is not necessarily Poisson compatible).

\begin{definition}
\label{def:Dequantization.StarProduct}
	The \defof{dequantization star product} $\star_{\varUpsilon}$ is the star product \eref{eq:StarProduct} with operators $B_{\varUpsilon, k} : \mathcal{A}_{0} \times \mathcal{A}_{0} \to \mathcal{A}_{0}$ such that for all $k \in \mathbb{N}$ and any dequantization expandable sections $A_1, A_2 \in \varGamma_{\varUpsilon} \subseteq \varGamma$, the expansion map intertwines $\star_{\varUpsilon}$ with the product on $\varGamma$, meaning
	\begin{align}
	\label{eq:Dequantization.StarProductCondition}
				\Sigma^{k}_{\varUpsilon}( A_1 A_2, \hbar )
		&= \sum_{j = 0}^{k}
				B_{\varUpsilon, j}\bigl(
					\Sigma^{k}_{\varUpsilon}( A_1, \hbar ),
					\Sigma^{k}_{\varUpsilon}( A_2, \hbar )
				\bigr)
				\hbar^{j}
		\mod \hbar^{k + 1}
		.
	\end{align}
\end{definition}

\begin{definition}
\label{eq:Dequantization.StrictDeformation}
	Given a continuous field of C*-algebras $\bigl( I, ( \mathfrak{A}_{\hbar} )_{\hbar \in I}, \varGamma \bigr)$, a dequantization $\varUpsilon$ is an \defof{infinite order strict deformation dequantization} if
	\begin{enumerate}
		\item the dequantization of any section $A \in \varGamma$,
			\begin{align}
			\label{eq:Dequantization.Field}
						\varUpsilon( A ) : \mathcal{M} \times I
				&\to \mathbb{C},
				&
						\varUpsilon( A )( x, \hbar )
				&\mapsto \varUpsilon_{\hbar}\bigl( A( \hbar ) \bigr)( x )
				,
			\end{align}
			is a continuous function $\varUpsilon( A ) \in \mathrm{C}( \mathcal{M} \times I, \mathbb{C} )$,
		\item the dequantization star product $\star_{\varUpsilon}$ exists, and
		\item the star product is Poisson compatible.
	\end{enumerate}
\end{definition}

The construction of a strict deformation quantization, a corresponding continuous field of C*-algebras $\bigl( I, ( \mathfrak{A}_{\hbar} )_{\hbar \in I}, \varGamma \bigr)$ and a corresponding strict deformation dequantization can be quite complicated in the general case. 
We consider strict deformation quantization associated to the (de)quantization maps that we obtain from geometric quantization in the case of a vector space in \Sref{sec:VectorSpace.GeometricQuantization}. 
Before coming to a general review of the necessary aspects of geometric quantization in the next section, let us introduce the Sorkin-Johnston state that we reconstruct later on. 

\subsection{The Sorkin-Johnston state as an algebraic state}
\label{subsec:SorkinJohnstonStateAlgebraic}
In algebraic quantum field theory, states are defined as functionals on a given *-algebra without requiring a Hilbert space. 
\begin{definition}
\label{def:State}
	A linear functional $\sigma : \mathcal{A} \to \mathbb{C}$ on an involutive algebra (*-algebra) $\mathcal{A}$ is a \defof{state} if and only if it is positive,
	\begin{align}
	\label{eq:State}
				\forall A
		&\in \mathcal{A}:
		&
				\sigma( A^* A )
		&\geq 0
		,
	\end{align}
	and has unit norm.
\end{definition}
For the definition of the Sorkin-Johnston state, in particular, consider the following class of states. 
\begin{definition}
\label{def:QuasiFreeState}
	Let $\mathfrak{W}_{\hbar}$ be the Weyl algebra for a real vector space $\mathcal{S}$ (see also the example in \Sref{subsec:FormalDeformationQuantization}).
	A state $\sigma$ on $\mathfrak{W}_{\hbar}$ is called \defof{quasi-free} (or Gaussian) if there exists a symmetric, bi-linear form $\gamma$ (called \defof{covariance} of the state) on $\mathcal{S}^*$ such that
	\begin{align}
	\label{eq:StateQuasiFree}
				\sigma\bigl( W_{\hbar}( \phi ) \bigr)
		&= \exp\left( - \frac{\hbar}{4} \gamma( \phi, \phi ) \right)
	\end{align}
	holds for the Weyl generator $W_{\hbar}( \phi )$ of every element $\phi \in \mathcal{S}^*$.
\end{definition}

The main argument for the Sorkin-Johnston state \cite{2010Johnston,2011Sorkin,2017Sorkin} is that there exists a Hermitian operator (as solution to the Sorkin-Johnston axioms) 
\begin{align}
\label{eq:SJTwoPointFunction.SJOperator}
			A_{\mathrm{SJ}}
	&= \frac{1}{2}
			\bigl( | E | + \mathrm{i} E \bigr)
\end{align}
that yields the ``positive eigenspace'' of the (Pauli-Jordan) operator $E$, and determines a two-point function \cite{2010Johnston}.
We call $A_{\mathrm{SJ}}$ the Sorkin-Johnston operator that acts on the complexified vector space $\mathcal{S}_{J}$.
By the one-to-one correspondence between two-point functions and quasi-free states, the bi-linear form $\eta$, see \eref{eq:VectorSpace.SymmetricForm}, determines the covariance of a quasi-free state.
\begin{definition}
	The \defof{Sorkin-Johnston state} $\sigma_{\mathrm{SJ}} : \mathfrak{W}_{\hbar} \to \mathbb{C}$ is the quasi-free (or Gaussian) state with a covariance given by the inverse of the symmetric, bi-linear form $\eta$ as defined in \eref{eq:VectorSpace.SymmetricForm}, so that for all $\phi \in \mathcal{S}^*$
	\begin{align}
	\label{eq:SJState}
				\sigma_{\mathrm{SJ}}\bigl( W_{\hbar}( \phi ) \bigr)
		&= \exp\left( - \frac{\hbar}{4} \eta^{-1}( \phi, \phi ) \right)
		.
	\end{align}
\end{definition}

More generally, for a quasi-free state with covariance $\gamma$, the bi-linear form $\eta_{\mathrm{G}} = \gamma^{-1}$ and the symplectic form $\omega$ satisfy the Cauchy-Schwarz inequality, $\forall v_1, v_2 \in \mathcal{S}$:
\begin{align}
\label{eq:QuasiFreeState.DominationCondition}
			\left| \omega( v_1, v_2 ) \right|^2
	&\leq \eta_{\mathrm{G}}( v_1, v_1 ) \eta_{\mathrm{G}}( v_2, v_2 )
	,
\end{align}
known as the domination condition \cite{2012FewsterVerch}.
\begin{figure}
	\centering
	\includegraphics[scale=\figscale]{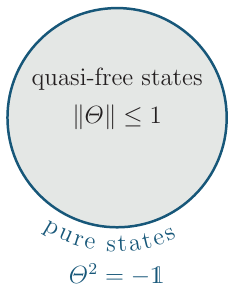}
	\caption{\label{fig:QuasifreeStates} Operator norm of the operator $\varTheta$ determined by the symplectic form and the inner product $\eta_{\mathrm{G}}$ as given in \eref{eq:QuasiFreeState.BilinearFormRelation}. For pure states, the operator norm lies on the boundary of this graphical representation and $\varTheta$ is a complex structure.}
\end{figure}
Given a relation between the bi-linear forms $\omega$ and $\eta_{\mathrm{G}}$ such that for all $v_1, v_2 \in \mathcal{S}$:
\begin{align}
\label{eq:QuasiFreeState.BilinearFormRelation}
			\omega( v_1, \varTheta v_2 )
	&= \eta_{\mathrm{G}}( v_1, v_2 )
	,
\end{align}
the domination condition implies that $\| \varTheta \| \leq 1$.
\Fref{fig:QuasifreeStates} illustrates this condition on the operator norm, where the norm is induced by the inner product.
For the proofs of the equivalences of the three statements, see \cite{2013DerezinskiGerard}.

For the structure that determines the Sorkin-Johnston state, the operator $\varTheta = J$ is a complex structure, meaning that $\varTheta^2 = -\mathbf{1}$, so that $\| \varTheta \| = 1$ and \eref{eq:QuasiFreeState.DominationCondition} is saturated.
This means that the Sorkin-Johnston state is \emph{pure} -- it cannot be written as a convex combination of two other states.
By the one-to-one correspondence between quasi-free states and two-point functions, the Sorkin-Johnston state corresponds to a two-point function determined by the Sorkin-Johnston operator $A_{\mathrm{SJ}}$.

\section{Aspects of geometric quantization}
\label{sec:GeometricQuantization}
For the geometric construction, we start with the given structure of the symplectic form $\omega$ and the inner product $\left\langle \cdot, \cdot \right\rangle$ as symmetric bi-linear form over $\mathcal{S}$ to construct a quantum algebra.
Our construction will yield the Sorkin-Johnston state without imposing the Sorkin-Johnston axioms in \Sref{subsec:DequantizationState}.
Throughout the construction, the quantization parameter $\hbar$ is kept explicit to eventually define a field of algebras over all $\hbar \in I$ and discuss the classical limit.

\subsection{Quantization bundle and polarizations}
\label{subsec:QuantizationBundle}
\begin{definition}
	Let $( \mathcal{M}, \omega )$ be a real, symplectic manifold.
	For some $\hbar$, a \defof{quantization bundle} is a Hermitian line bundle $\mathcal{L}_{\hbar} \to \mathcal{M}$ with connection $\nabla_{\hbar}$ that preserves the inner product, such that its curvature $\operatorname{curv}( \nabla_{\hbar} )$ is proportional to the symplectic form,
	\begin{align}
	\label{eq:QBundle.Curvature}
				\operatorname{curv}( \nabla_{\hbar} )
		&= - \frac{\mathrm{i}}{\hbar} \omega
		.
	\end{align}
\end{definition}
Given a symplectic manifold, a quantization bundle does not necessarily exist and is not necessarily unique.
A quantization bundle for $(\mathcal M,\omega)$ exists if and only if the cohomology class of $\omega / 2 \pi \hbar$ in $H^2( \mathcal{M}, \mathbb{R} )$ is integral. This is known as the \emph{prequantization (or integrality) condition}, see \cite[Sec.~3]{2005AliEnglis}.

For geometric quantization, it is furthermore necessary to find a ``physical'' Hilbert space $\mathcal{H}_{\hbar}$ as a subspace of square-integrable sections $\mathrm{L}^2( \mathcal{M}, \mathcal{L}_{\hbar} )$ (or valued in $\mathcal{L}_{\hbar}$ tensored with some vector bundle).
In some cases, the space $\mathcal{H}_{\hbar}$ is determined by a polarization \cite{2005AliEnglis}.

\begin{definition}
\label{def:QBundle.Polarization}
	Let $( \mathcal{M}, \omega )$ be an $2 N$-dimensional real symplectic manifold and $\mathcal{L}_{\hbar} \to \mathcal{M}$ a quantization bundle.
	A (complex) \defof{polarization} is a subbundle $P \subset ( \mathrm{T} \mathcal{M} )^{\mathbb{C}}$ that is involutive, $X, Y \in \varGamma( P ) \implies [ X, Y ] \in \varGamma( P )$, and maximally isotropic (Lagrangian), $\forall X, Y \in \varGamma( P ): \omega( X, Y ) = 0$ and $\forall x \in \mathcal{M}: \dim_{\mathbb{C}} P_{x} = N$.
	We say that a section $\psi \in \varGamma( \mathcal{M}, \mathcal{L}_{\hbar} )$ is \defof{polarized} if $\forall X \in \varGamma( P ): \nabla_{\hbar, X} \psi = 0$.
	The \defof{physical Hilbert space} $\mathcal{H}_{\hbar}$ is constructed from polarized sections of $\mathcal{L}_{\hbar}$.
\end{definition}

For a K\"ahler manifold (a symplectic manifold with a compatible complex structure, $J$), the \defof{K\"ahler polarization} is the subbundle on which $J$ has eigenvalue $-\mathrm{i}$, and the polarized sections are precisely the holomorphic sections.
Compact K\"ahler manifolds have been studied before, see \cite{1994BordemannMeinrenkenSchlichenmaier,2001KarabegovSchlichenmaier,2010Schlichenmaier}.

For the more general case of a symplectic manifold without pre-defined complex structure, we will consider an alternative construction of the physical Hilbert space from the spectrum of a Laplace operator in \Sref{sec:VectorSpace.GeometricQuantization}.

\subsection{(Berezin-)Toeplitz quantization and dequantization}
\label{subsec:BerezinToeplitz}
For a given symplectic manifold $( \mathcal{M}, \omega )$, suppose that a Hilbert subspace $\mathcal{H}_{\hbar} \subset \mathrm{L}^2( \mathcal{M}, \mathcal{L}_{\hbar} )$ has been constructed.
Let
\begin{align}
\label{eq:BTQuantization.Projector}
			\varPi_{\hbar} : \mathrm{L}^2( \mathcal{M}, \mathcal{L}_{\hbar} )
	&\to \mathcal{H}_{\hbar}
\end{align}
be the projector to this Hilbert space.
With the given Hilbert space $\mathcal{H}_{\hbar}$, we also obtain a quantization map \cite{1981deMonvelGuillemin}.

\begin{definition}
\label{def:BTQuantization}
	The \defof{(Berezin-)Toeplitz quantization} map $T_{\hbar}$ assigns a bounded operator on the Hilbert space $\mathcal{H}_{\hbar}$ to every classical observable,
	\begin{align}
	\label{eq:BTQuantization}
				T_{\hbar} : \mathcal{A}_{0}
		&\to \mathcal{B}( \mathcal{H}_{\hbar} )
		,
	\end{align}
	using the projection $\varPi_{\hbar}$ from the space of square-integrable sections $\mathrm{L}^2( \mathcal{M}, \mathcal{L}_{\hbar} )$ such that
	\begin{align}
	\label{eq:BTQuantization.Operator}
				\forall \psi
		&\in \mathcal{H}_{\hbar}:
		&
				T_{\hbar}( f ) \psi
		&= \varPi_{\hbar}( f \psi )
		.
	\end{align}
\end{definition}

For each $\hbar \in I$, we also choose an algebra $\mathfrak{A}_{\hbar} \subseteq \mathcal{B}( \mathcal{H}_{\hbar} )$ such that it contains the image of $T_{\hbar}$.
The Toeplitz quantization map $T_{\hbar}$ is linear and respects involution.
Its domain actually extends to all bounded functions $\mathrm{C}_\mathrm{b}( \mathcal{M}, \mathbb{C} )$ so that $1 \in \mathrm{C}_\mathrm{b}( \mathcal{M}, \mathbb{C} )$ is mapped to $\mathbf{1} \in \mathcal{B}( \mathcal{H}_{\hbar} )$.
However, it will be easier to use the C*-algebra of compact operators $\mathfrak{A}_{\hbar} = \mathcal{K}( \mathcal{H}_{\hbar} )$ in the construction of a continuous field of C*-algebras. Note that $\mathfrak{A}_{\hbar}$ coincides with $\mathcal{B}( \mathcal{H}_{\hbar} )$ if $\dim \mathcal{H}_{\hbar} < \infty$.
Furthermore, it will also be necessary to restrict to a dense subalgebra $\mathcal{A}_{0} \subset \mathfrak{A}_{0}$ for the construction of formal deformation quantizations (star products).

If the Toeplitz operators of compactly supported functions $\mathrm{C}_\mathrm{c}( \mathcal{M}, \mathbb{C} )$ on the physical Hilbert space $\mathcal{H}_{\hbar}$ are of trace-class, we define a measure $\mu_{\hbar}$ such that
\begin{align}
\label{eq:BTDequantization.Measure}
			\operatorname{Tr}\bigl( T_{\hbar}( f ) \bigr)
	&= \int_{\mathcal{M}} f \,\mathrm{d}\mu_{\hbar}
\end{align}
holds for all $f \in \mathrm{C}_\mathrm{c}( \mathcal{M}, \mathbb{C} )$.
When such a measure exists, we have an adjoint operation to Toeplitz quantization.

\begin{definition}
\label{def:BTDequantization}
	Suppose the measure $\mu_{\hbar}$ determined by \eref{eq:BTDequantization.Measure} exists.
	The \defof{(Berezin)-Toeplitz dequantization} is a family of linear maps
	\begin{align}
	\label{eq:BTDequantization}
				\varXi_{\hbar} : \mathfrak{A}_{\hbar}
		&\to \mathfrak{A}_{0}
		,
	\end{align}
	such that for all complex-valued, compactly supported functions $f \in \mathrm{C}_\mathrm{c}( \mathcal{M}, \mathbb{C} )$ and all operators $A_{\hbar} \in \mathfrak{A}_{\hbar}$
	\begin{align}
	\label{eq:BTDequantization.Dual}
				\operatorname{Tr}( A_{\hbar} T_{\hbar}( f ) )
		&= \int_{\mathcal{M}} \varXi_{\hbar}( A_{\hbar} ) f \,\mathrm{d}\mu_{\hbar}
		.
	\end{align}
\end{definition}
Consider the case of a symplectic manifold with a physical Hilbert space $\mathcal{H}_{\hbar}$ such that Toeplitz dequantization exists.
If the algebras $\mathfrak{A}_{\hbar}$ are unital, then Toeplitz dequantization preserves the unit, $\varXi_{\hbar}( \mathbf{1} ) = 1$ since the measure is normalized,
\begin{align}
\label{eq:BTDequantization.Dual.Unital}
			\forall f
	&\in \mathrm{C}_\mathrm{c}( \mathcal{M}, \mathbb{C} ):
	&
			\int_{\mathcal{S}} \varXi_{\hbar}( \mathbf{1} ) f \,\mathrm{d}\mu_{\hbar}
	&= \int_{\mathcal{S}} f \,\mathrm{d}\mu_{\hbar}
	.
\end{align}
Applying dequantization to a Toeplitz operator $T_{\hbar}( f )$ yields a ``smearing'' of the original function.
For more details on Berezin-Toeplitz dequantization, see also \cite{2020IoosKaminkerPolterovichShmoish}.

\begin{definition}
	The \defof{Berezin transform} of a classical observable $f \in \mathcal{A}_{0}$ over the symplectic manifold $( \mathcal{M}, \omega )$ is the dequantization of its Toeplitz operator, $( \varXi_{\hbar} \after T_{\hbar} )( f )$.
\end{definition}
A function $f \in \mathcal{A}_{0}$ is sometimes referred to as the \emph{contravariant} or \emph{lower} symbol of the Toeplitz operator $T_{\hbar}( f )$, while the Berezin transform $( \varXi_{\hbar} \after T_{\hbar} )( f )$ is also called the \emph{covariant} or \emph{upper} symbol of $T_{\hbar}( f )$ \cite{1972Berezin}.

\subsection{Laplacians on the quantization bundle}
\label{sec:QuantizationRiemannManifold}
Given a symplectic manifold with Riemannian metric, we want to identify the physical Hilbert space as a subspace of quantization bundle sections that correspond to the lowest part of the spectrum of the Laplacian defined with the metric.
In this section, we review some general arguments for symplectic manifolds to motivate a generalisation of our results in \Sref{sec:VectorSpace.GeometricQuantization}.

\begin{definition}
\label{def:BochnerLaplacian}
	Let $( \mathcal{M}, \omega, g )$ be a symplectic manifold with Riemannian metric, $\mathcal{L}_{\hbar} \to \mathcal{M}$ be a quantization bundle for some $\hbar \in I_*$.
	The \defof{Bochner Laplacian} $\Laplace_{\hbar}$ is an unbounded operator on square-integrable, smooth sections of the bundle. 
	It is determined by the connection $\nabla_{\hbar}$ and metric $g$,
	\begin{align}
	\label{eq:QBundle.BochnerLaplacian}
				\Laplace_{\hbar}
		&= \nabla_{\hbar}^* \nabla_{\hbar}
		.
	\end{align}
\end{definition}

In the case of a $2 N$-dimensional K\"ahler manifold, let $\nabla_{i}$ denote the holomorphic and $\nabla_{\jb}$ the anti-holomorphic components of the connection $\nabla_{\hbar}$, with $i \in [ 1, N ]$, $\jb \in [ \bar{1}, \bar{N} ]$.
There is another, naturally defined Laplace operator, the Kodaira Laplacian
\begin{align}
\label{eq:QBundle.KodairaLaplacian}
			\Laplace^{\mathrm{K}}_{\hbar}
	&= - g^{i \jb} \nabla_{i} \nabla_{\jb}
\end{align}
using the summation convention.
The Kodaira Laplacian is related to the Bochner Laplacian,
\begin{align}
\label{eq:QBundle.LaplacianRelation}
			2 \Laplace^{\mathrm{K}}_{\hbar}
	&= \Laplace_{\hbar}
		- \frac{N}{\hbar}
	.
\end{align}
With the K\"ahler polarization, the physical Hilbert space is constructed from the space of holomorphically polarized sections with respect to the complex structure of the K\"ahler manifold.
The kernel of the Kodaira Laplacian \eref{eq:QBundle.KodairaLaplacian} contains the space of holomorphic sections.
In fact, the  kernel is precisely the space of holomorphic sections since the holomorphic components $\nabla_{i}$ are adjoint to the anti-holomorphic components $\nabla_{\ib}$.
The Kodaira  Laplacian is positive, so the space of holomorphic sections is the eigenspace of the Bochner Laplacian corresponding to the lowest eigenvalue $\frac{N}{\hbar}$, see \eref{eq:QBundle.LaplacianRelation}.
Thus, the physical Hilbert space is equivalently determined from the spectrum of the Bochner Laplacian.

In \cite{1988GuilleminUribe}, it was shown how to use a renormalized Bochner Laplacian for a natural generalization to almost K\"ahler manifolds (with a non-integrable, almost complex structure).
The renormalized Bochner Laplacian is a generalization of the expression on the right hand side of \eref{eq:QBundle.LaplacianRelation} and coincides with $2 \Laplace^{\mathrm{K}}_{\hbar}$ in the K\"ahler case, see also \cite{1996BorthwickUribe}.
A choice of a physical Hilbert space is again given by the eigenspace corresponding to the lowest part of the spectrum, even though the lowest part does not have to be a single eigenvalue anymore.

A further generalization starts with a symplectic manifold with Riemannian metric without pre-defined complex structure, but for bounded geometry at infinity, see \cite{2008MaMarinescu-Bergman,2008MaMarinescu-Toeplitz,2019KordyukovMaMarinescu}.
For this, consider a $2 N$-dimensional, compact, real, symplectic manifold $( \mathcal{M}, \omega )$ with quantization bundle $\mathcal{L}_{\hbar} \to \mathcal{M}$ and Riemannian metric $g$.
There exists an anti-self-adjoint linear map $E : \mathrm{T} \mathcal{M} \to \mathrm{T} \mathcal{M}$ such that for all $v_1, v_2 \in \mathrm{T} \mathcal{M}$
\begin{align}
\label{eq:Manifold.StructureRelation}
			\omega( v_1, v_2 )
	&= g( v_1, E^{-1} v_2 )
	.
\end{align}
There exists an almost complex structure $J : \mathrm{T} \mathcal{M} \to \mathrm{T} \mathcal{M}$ such that $g( J v_1, J v_2 ) = g( v_1, v_2 )$ and $\omega( J v_1, J v_2 ) = \omega( v_1, v_2 )$ for all $v_1, v_2 \in \mathrm{T} \mathcal{M}$.
We define a new metric $\eta$ such that for all $v_1, v_2 \in \mathrm{T} \mathcal{M}$
\begin{align}
\label{eq:Manifold.AlmostKaehlerMetric}
			\eta( v_1, v_2 )
	&:= \omega( v_1, J v_2 )
	.
\end{align}
The almost complex structure commutes with $E^{-1}$, so
\begin{align}
\label{eq:Manifold.AlmostComplexStructure}
			J
	&= - E^{-1} | E |
	.
\end{align}
At every point $x \in \mathcal{M}$, the operator $E^{-1}_{x}$ is an endomorphism and we denote half the trace of $| E_{x} |^{-1}$ as
\begin{align}
\label{eq:Manifold.StructureRelationTrace}
			\lambda( x )
	&:= \frac{1}{2} \operatorname{tr}\bigl( J_{x} E^{-1}_{x} \bigr)
	= \frac{1}{2} \operatorname{tr} | E_{x} |^{-1}
	.
\end{align}
Note that in the special case of a K\"ahler manifold with a K\"ahler metric so that $E_{x} = \operatorname{id}$, this trace is $N$, half the real dimension of $\mathcal{M}$.

It was shown that the trace $\lambda( x )$ is positive for all $x \in \mathcal{M}$.
A renormalized Bochner Laplacian $\Laplace_{\hbar, \varPhi}$ is then defined with $\lambda$ and a smooth Hermitian section $\varPhi$ on a tensor product of the quantization line bundle with a vector bundle, see \cite{2002MaMarinescu}.
They have shown --- using $\mathrm{Spin}^{\mathbb{C}}$ Dirac operators --- that there exist two positive constants $\kappa$ and $\mu$ that are independent of $\hbar$, such that the spectrum of the renormalized Bochner Laplacian fulfills
\begin{align}
\label{eq:QBundle.BochnerLaplacianRenormalizedSpectrum}
			\operatorname{spec}( \Laplace_{\hbar, \varPhi} )
	&\subset [ -\kappa, \kappa ]
		\cup \left[ \frac{2 \mu}{\hbar} - \kappa, \infty \right)
	.
\end{align}
Given this spectrum condition for the renormalized Bochner Laplacian on a symplectic manifold with Riemannian metric $( \mathcal{M}, \omega, g )$, the physical Hilbert space $\mathcal{H}_{\hbar} \subset \mathrm{L}^2( \mathcal{M}, \mathcal{L}_{\hbar} )$ is spanned by the sections corresponding to the lower part of the spectrum i.e., the part contained in $[ -\kappa, \kappa ]$.

\section{Geometric quantization for a symplectic vector space with inner product}
\label{sec:VectorSpace.GeometricQuantization}
For a symplectic vector space with inner product $( \mathcal{S}, \omega, \left\langle \cdot, \cdot \right\rangle )$ as described in \Sref{subsec:ScalarFieldsCausalSets}, we now use the idea of the geometric construction to derive a physical Hilbert space $\mathcal{H}_{\hbar}$.
The inner product corresponds to the metric $g$ and we use a basis of holomorphic and anti-holomorphic vectors to express components with indices raised and lowered by $g$.
This choice of a complex basis will allow us to write the sections of the Hilbert space as holomorphic sections with respect to the complex structure $J$ given in \eref{eq:Manifold.AlmostComplexStructure}.
Further details on this construction are also given in \cite[Sec.~4.1.6]{2007MaMarinescu} and \cite[Sec.~1.4]{2008MaMarinescu-Bergman}.

\subsection{The Bochner Laplacian and its spectrum}
\label{subsec:VectorSpace.BochnerLaplacian}
On the vector space $\mathcal{S}$, consider an exact symplectic form $\omega = -\mathrm{d}{\theta}$ such that we have a trivial line bundle $\mathcal{L}_{\hbar} := \mathcal{S} \times \mathbb{C}$ with non-trivial connection parametrized by $\hbar$,
\begin{align}
\label{eq:QBundle.Connection}
			\nabla_{\hbar}
	&= \mathrm{d}
		+ \frac{\mathrm{i}}{\hbar} \theta
	.
\end{align}
With the complex structure $J$ given as in \eref{eq:Manifold.AlmostComplexStructure}, we turn the real vector space $\mathcal{S}$ into a complex vector space $\mathcal{S}_{J}$ by the assignment \eref{eq:VectorSpace.ComplexStructure}.
The operator $\nabla_{\hbar}$ increases the total degree $p + q$ of complex differential forms $\Omega^{p, q}$ by 1.
We define operators $\mathcal{D}^{+}_{\hbar} : \Omega^{p, q} \to \Omega^{p + 1, q}$ raising the holomorphic degree and $\conj{\mathcal{D}}{}^{+}_{\hbar} : \Omega^{p, q} \to \Omega^{p, q + 1}$ raising the anti-holomorphic degree such that
\begin{align}
\label{eq:QBundle.ConnectionSplit}
			\nabla_{\hbar}
	&= \mathcal{D}^{+}_{\hbar} + \conj{\mathcal{D}}{}^{+}_{\hbar}
	.
\end{align}
Use the Hodge dual operator ${\ast} : \Omega^{p, q} \to \Omega^{N - q, N - p}$ to define the adjoint operators 
\begin{subequations}
\label{eq:QBundle.ConnectionDualSplit}
\begin{align}
			\mathcal{D}^{-}_{\hbar}
	&:= - {\ast} \conj{\mathcal{D}}{}^{+}_{\hbar} {\ast}
	,
	\allowdisplaybreaks[3]\\
			\conj{\mathcal{D}}{}^{-}_{\hbar}
	&:= - {\ast} \mathcal{D}^{+}_{\hbar} {\ast}
	.
\end{align}
\end{subequations}

Thus the Bochner Laplacian is given as 
\begin{align}
\label{eq:QBundle.BochnerLaplacianSplit}
			\Laplace_{\hbar}
	&= \mathcal{D}^{-}_{\hbar} \mathcal{D}^{+}_{\hbar}
			+ \mathcal{D}^{-}_{\hbar} \conj{\mathcal{D}}{}^{+}_{\hbar}
			+ \conj{\mathcal{D}}{}^{-}_{\hbar} \mathcal{D}^{+}_{\hbar}
			+ \conj{\mathcal{D}}{}^{-}_{\hbar} \conj{\mathcal{D}}{}^{+}_{\hbar}
	.
\end{align}
Note that we are want to act on $( 0, 0 )$-forms, for which the two middle terms vanish. 

For the following computation, we choose complex coordinates $z^{i} = x^{i} + \mathrm{i} y^{i}$ (with indices $i \in [ 1, N ]$, $\ib \in [ \bar{1}, \bar{N} ]$) in which $| E |^{-1}$ is diagonal with diagonal components $\vartheta_{i}$.
The indices are raised with the inverse metric $g^{\ib i}$ derived from the inner product on $\mathcal{S}$.
Raising an index also changes it from holomorphic to anti-holomorphic and vice versa.
For a slightly compacter notation, we omit the $\hbar$ subscript whenever the connection $\nabla_{\hbar}$ is expressed in coordinates.

The difference of the first and last operator pair in \eref{eq:QBundle.BochnerLaplacianSplit} is 
\begin{subequations}
\label{eq:QBundle.CurvatureContraction}
\begin{align}
			\mathcal{D}^{-}_{\hbar} \mathcal{D}^{+}_{\hbar}
		- \conj{\mathcal{D}}{}^{-}_{\hbar} \conj{\mathcal{D}}{}^{+}_{\hbar}
	&= g^{\ib j} \comm{\nabla_{j}}{\nabla_{\ib}}
	\allowdisplaybreaks[3]\\
	&= - \frac{\mathrm{i}}{\hbar} g^{\ib j} \omega_{j \ib}
	=: \lambda_{\hbar}
	.
\end{align}
\end{subequations}
We use this identity to replace the first operator pair of the Bochner Laplacian \eref{eq:QBundle.BochnerLaplacian} by the anti-holomorphic operators $\conj{\mathcal{D}}{}^{\pm}_{\hbar}$ along with the positive shift constant $\lambda_{\hbar}$, such that we obtain for $( 0, 0 )$-forms
\begin{align}
\label{eq:QBundle.BochnerLaplacianDDerivMP}
			\Laplace_{\hbar}
	&= 2 \conj{\mathcal{D}}{}^{-}_{\hbar} \conj{\mathcal{D}}{}^{+}_{\hbar}
		+ \lambda_{\hbar} \mathbf{1}
	,
\end{align}
which is relating the Bochner and Kodaira Laplacians as in the case of a K\"ahler space \eref{eq:QBundle.LaplacianRelation}, see also \cite[Eq.~1.4.31]{2007MaMarinescu}. 
The constant $\lambda_{\hbar}$ is (up to the quantization parameter $\hbar$) half the trace of $| E |^{-1}$,
\begin{align}
\label{eq:QBundle.GroundStateEigenvalue}
			\lambda_{\hbar}
	&= \frac{1}{2 \hbar} \operatorname{tr} | E |^{-1}
	.
\end{align}
In our choice of complex coordinates such that $| E |^{-1}$ is diagonal, this constant is half the sum over all the diagonal components divided by $\hbar$.

Combine the operators $\conj{\mathcal{D}}{}^{\pm}_{\hbar} : \Omega^{0, q} \to \Omega^{0, q \pm 1}$ that increase and decrease the anti-holomorphic degree to a self-adjoint operator,
\begin{align}
\label{eq:QBundle.DDerivSelfAdjoint}
			\conj{\mathcal{D}}_{\hbar}
	&:= \conj{\mathcal{D}}{}^{+}_{\hbar} + \conj{\mathcal{D}}{}^{-}_{\hbar}
	.
\end{align}
So the Laplacian acting on $( 0, 0 )$-forms $\psi \in \Omega^{0, 0}$ becomes
\begin{align}
\label{eq:QBundle.BochnerLaplacianDDerivSquare}
			\Laplace_{\hbar} \psi
	&= 2 \conj{\mathcal{D}}_{\hbar}^2 \psi
		+ \lambda_{\hbar} \psi
	= - 2 g^{i \jb} \nabla_{i} \nabla_{\jb} \psi + \lambda_{\hbar} \psi
	.
\end{align}
Since by construction the Laplacian and the operator $\conj{\mathcal{D}}_{\hbar}^2$ are self-adjoint and positive, $\lambda_{\hbar}$ is the lower bound on the spectrum of the Laplacian.
We express the components of the symmetric bi-linear form $g^{i \jb}$ (corresponding to the inner product $\left\langle \cdot, \cdot \right\rangle$) in terms of the diagonal components $\vartheta_{i}$ of $| E |^{-1}$,
\begin{subequations}
\label{eq:VectorSpace.MetricComponents}
\begin{align}
			\bigl( g_{i \jb} \bigr)_{i \in [ 1, N ], \jb \in [ \bar{1}, \bar{N} ]}
	&= \operatorname{diag}\Bigl(
				\frac{1}{\vartheta_{1}},
				\frac{1}{\vartheta_{2}},
				\dotso,
				\frac{1}{\vartheta_{N}}
			\Bigr)
	,
	\allowdisplaybreaks[3]\\
			\bigl( g^{i \jb} \bigr)_{i \in [ 1, N ], \jb \in [ \bar{1}, \bar{N} ]}
	&= \operatorname{diag}\bigl(
				\vartheta_{1},
				\vartheta_{2},
				\dotso,
				\vartheta_{N}
			\bigr)
	.
\end{align}
\end{subequations}
The components of the covariant derivative fulfill the commutation relations (see also \cite[Eq.~4.1.75]{2007MaMarinescu} and \cite[Eq.~1.86]{2008MaMarinescu-Bergman})
\begin{subequations}
\label{eq:QBundle.ConnectionCommutators}
\begin{align}
			\comm{\nabla_{i}}{\nabla_{j}}
	&= 0
	,
	\allowdisplaybreaks[3]\\
			\comm{\nabla_{i}}{\nabla_{\jb}}
	&= - \frac{\mathrm{i}}{\hbar} \omega_{i \jb}
	= \frac{1}{\hbar} \delta_{i \jb}
	,
	\allowdisplaybreaks[3]\\
			\comm{\nabla_{\ib}}{\nabla_{\jb}}
	&= 0
	.
\end{align}
\end{subequations}
With the component representation as given here, we find the full spectrum of the Bochner Laplacian, see \cite[Thm.~4.1.20]{2007MaMarinescu} and \cite[Thm.~1.15]{2008MaMarinescu-Bergman}).

\begin{theorem}
	Let $\Laplace_{\hbar}$ be the Bochner Laplacian for square-integrable sections of the quantization bundle $\mathcal{L}_{\hbar} \to \mathcal{S}$ over a real $2 N$-dimensional symplectic vector space $( \mathcal{S}, \omega, \left\langle \cdot, \cdot \right\rangle )$ with a non-degenerate symplectic form $\omega$ and an inner product $\left\langle \cdot, \cdot \right\rangle$.
	Let $\vartheta_{i}$ be the diagonal components as given in \eref{eq:VectorSpace.MetricComponents}.
	The spectrum of the Laplacian is given by
	\begin{align}
	\label{eq:QBundle.LaplacianSpectrum}
				\operatorname{spec}( \Laplace_{\hbar} )
		&= \left\{
					\frac{1}{\hbar}
					\sum_{i = 1}^{N}
						( 2 n_{i} + 1 )
						\vartheta_{i}
				\,\middle\vert\,
					n_{i} \in \mathbb{N}
				\right\}
		.
	\end{align}
\end{theorem}
\begin{figure}
	\centering
	\includegraphics[scale=\figscale]{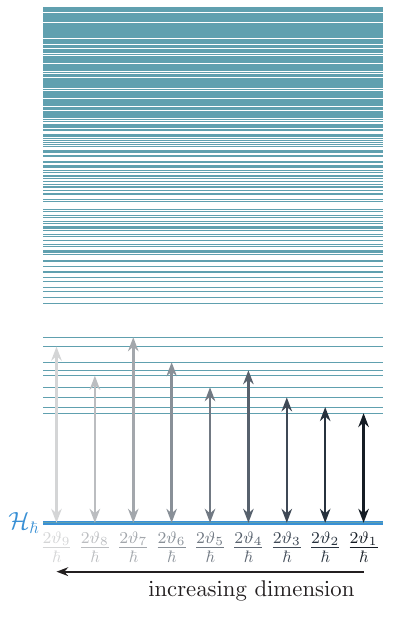}
	\caption{\label{fig:LaplaceLevels} Illustration of the spectrum of the Bochner Laplacian $\Laplace_{\hbar} = \nabla_{\hbar}^* \nabla_{\hbar}$ for a symplectic vector space with 18 dimensions ($N = 9$) and diagonal metric components $\vartheta_{i}$. The Hilbert space $\mathcal{H}_{\hbar}$ is constructed from the section \eref{eq:QBundle.HolomSection} corresponding to the lowest spectral value (cyan) is the solution.}
\end{figure}
\Fref{fig:LaplaceLevels} depicts the spectrum, which gets denser towards infinity.
Note that the spectrum \eref{eq:QBundle.LaplacianSpectrum} does not depend on the complex structure $J$.
We use the space of eigen\-sections at the lowest spectral value as the physical Hilbert space $\mathcal{H}_{\hbar}$ on which bounded operators act as quantized observables.
The complex structure $J$ is determined by the operator $E^{-1}$ in \eref{eq:Manifold.AlmostComplexStructure} such that the eigenspace of the lowest spectral value corresponds to the holomorphic sections \eref{eq:QBundle.HolomSection}.

\subsection{The physical Hilbert space}
\label{subsec:VectorSpace.HilbertSpace}

The canonical and real-valued form of the symplectic potential corresponding to the chosen complex coordinates is
\begin{align}
\label{eq:VectorSpace.SymmetricPotential}
			\theta
	&= \frac{\mathrm{i}}{2} \delta_{i \jb}
			\left(
				\conj{z}^{\jb} \mathrm{d}{z^{i}}
			- z^{i} \mathrm{d}{\conj{z}^{\jb}}
			\right)
	,
\end{align}
and we write $| z |^2 = \delta_{i \ib} z^{i} \conj{z}^{\ib}$. 
In this gauge, any holomorphic section $\psi$ has an arbitrary, smooth, holomorphic function $\alpha$ as amplitude, so that for all $z \in \mathcal{S}$
\begin{align}
\label{eq:QBundle.HolomSection}
			\psi( z )
	&= \frac{\alpha( z )}{\sqrt{2 \pi \hbar}^{N}}
			\exp\left(
			- \frac{1}{2 \hbar} | z |^2
			\right)
	,
\end{align}
and it is a solution of the differential equation
\begin{align}
\label{eq:QBundle.DDerivMP00Form}
			\conj{\mathcal{D}}{}^{-}_{\hbar} \conj{\mathcal{D}}{}^{+}_{\hbar} \psi
	= - g^{i \jb} \nabla_{i} \nabla_{\jb} \psi
	&= 0
	.
\end{align}
The physical Hilbert space $\mathcal{H}_{\hbar} \subset \mathrm{L}^2( \mathcal{S}, \mathcal{L}_{\hbar} )$ is spanned by the sections corresponding to the lowest part of the spectrum, these are the \emph{holomorphic} sections of the quantization bundle $\mathcal{L}_{\hbar}$ that take the form \eref{eq:QBundle.HolomSection} in the complex coordinates.

The Hilbert space has the inner product
\begin{subequations}
\begin{align}
\label{eq:HilbertSpace.InnerProd}
			\braket{\psi_{1}}{\psi_{2}}_{\hbar}
	&= \int_{\mathcal{S}} \conj{\psi_{1}} \psi_{2} \,\mathrm{dvol}
	,
	\allowdisplaybreaks[3]\\\label{eq:VectorSpace.SymplecticVolumeForm}
			\mathrm{dvol}
	&= ( -1 )^{\frac{1}{2} N ( N - 1 )} \frac{1}{N!}
			\bigwedge_{i = 1}^{N} \omega
	.
\end{align}
\end{subequations}
Let $\mathfrak{A}_{0}$ be the C*-algebra $\mathrm{C}_0( \mathcal{S}, \mathbb{C} )$ and $\mathfrak{A}_{\hbar}$ be the C*-algebra of compact operators $\mathcal{K}( \mathcal{H}_{\hbar} )$.
Define Toeplitz quantization $T_{\hbar} : \mathrm{C}^\infty_\mathrm{S}( \mathcal{S}, \mathbb{C} ) \to \mathfrak{A}_{\hbar}$ as in \Defref{def:BTQuantization} with the projector $\varPi_{\hbar}$ as defined in \eref{eq:BTQuantization.Projector}.
Note that Toeplitz quantization actually extends to a map from bounded functions to bounded operators, $\mathrm{C}_\mathrm{b}( \mathcal{S}, \mathbb{C} ) \to \mathcal{B}( \mathcal{H}_{\hbar} )$, but we will restrict to Schwartz functions to construct a continuous field of C*-algebras later on.

For any two holomorphic sections $\psi_{1,2}$ as in \eref{eq:QBundle.HolomSection} with smooth, holomorphic functions $\alpha_{1,2}$ as amplitudes, the inner product reads
\begin{align}
\label{eq:HilbertSpace.InnerProdHolomSections}
			\braket{\psi_1}{\psi_2}_{\hbar}
	&= \frac{1}{( 2 \pi \hbar )^{N}}
			\int_{\mathcal{S}}
				\conj{\alpha_1}
				\alpha_2
				\exp\left(
				- \frac{1}{\hbar} | z |^2
				\right)
			\,\mathrm{dvol}
	.
\end{align}
Use the section basis given in ket-notation for any $n_{1}, \dots, n_{N} \in \mathbb{N}$, similar to \cite[Eq.~4.1.83]{2007MaMarinescu} and \cite[Eq.~1.89]{2008MaMarinescu-Bergman}, 
\begin{align}
\label{eq:HilbertSpace.Basis}
			\ket{n_{1}, \dots, n_{N}}_{\hbar}
	&:= \ket{\frac{1}{\sqrt{2 \pi \hbar}^{N}}
			\left(
			\prod_{i = 1}^{N}
				\frac{1}{\sqrt{n_{i}!}}
				\left( \frac{z^{i}}{\sqrt{\hbar}} \right)^{n_{i}}
			\right)
			\exp\left(
			- \frac{1}{2 \hbar} | z |^2
			\right)
			}_{\hbar}
	.
\end{align}
This basis is orthonormal,
\begin{align}
\label{eq:HilbertSpace.BasisOrthonormality}
			\braket{m_{1}, \dots, m_{N}}{n_{1}, \dots, n_{N}}_{\hbar}
	&= \prod_{i = 1}^{N} \delta_{m_{i} n_{i}}
	.
\end{align}
We may now define unbounded ladder operators (in the summation convention), see also \cite[Eq.~4.1.74]{2007MaMarinescu} and \cite[Eq.~1.85]{2008MaMarinescu-Bergman}, 
\begin{subequations}
\label{eq:HilbertSpace.CreationAnnihilationOperators}
\begin{align}
			a^{+}_{\ib}
	&:= \frac{1}{\sqrt{\hbar}} \delta_{\ib i} z^{i}
		- \sqrt{\hbar} \nabla_{\ib}
	\\
			a^{-}_{i}
	&:= \frac{1}{\sqrt{\hbar}} \delta_{i \ib} \conj{z}^{\ib}
		+ \sqrt{\hbar} \nabla_{i}
	,
\end{align}
\end{subequations}
which are adjoint to each other for each index pair $i = \ib$.
Using the commutators of the quantization bundle connection \eref{eq:QBundle.ConnectionCommutators}, we find the commutators for the ladder operators to be those known from an $N$-dimensional quantum mechanical harmonic oscillator,
\begin{subequations}
\label{eq:HilbertSpace.CreationAnnihilationCommutators}
\begin{align}
			\comm{a^{\pm}_{i}}{a^{\pm}_{j}}
	&= 0
	,
	\allowdisplaybreaks[3]\\
			\comm{a^{-}_{i}}{a^{+}_{\jb}}
	&= \delta_{i \jb} \mathbf{1}
	.
\end{align}
\end{subequations}
The action of the ladder operators on the Hilbert space basis yields
\begin{subequations}
\begin{align}
\label{eq:HilbertSpace.BasisCreation}
			a^{+}_{i} \ket{n_{1}, \dots, n_{i}, \dots, n_{N}}_{\hbar}
	&= \sqrt{n_{i} + 1} \ket{n_{1}, \dots, n_{i} + 1, \dots, n_{N}}_{\hbar}
	,
	\allowdisplaybreaks[3]\\\label{eq:HilbertSpace.BasisAnnihilation}
			a^{-}_{i} \ket{n_{1}, \dots, n_{i}, \dots, n_{N}}_{\hbar}
	&= \sqrt{n_{i}} \ket{n_{1}, \dots, n_{i} - 1, \dots, n_{N}}_{\hbar}
	.
\end{align}
\end{subequations}
The heuristic relation between the Toeplitz operators of the (unbounded) coordinate functions $z^{i}$ and $\conj{z}^{\ib}$ with the ladder operators as well as examples for anti-normal ordering of Toeplitz quantization are given in \ref{sec:AppendixNormalAntiNormalOrdering}.

Before we come to the construction of a continuous field of C*-algebras from the family of Hilbert spaces parametrized in $\hbar$, we use dequantization to determine the explicit form of the Berezin transform in the following.
Dequantization is the operation that leads us to the definition of a state -- the Sorkin-Johnston state.

\subsection{Dequantization and the Berezin transform}
\label{subsec:VectorSpace.Dequantization}
We defined the adjoint map $\varXi_{\hbar} : \mathfrak{A}_{\hbar} \to \mathfrak{A}_{0}$ to Toeplitz quantization by the relation \eref{eq:BTDequantization.Dual}, which becomes
\begin{align}
\label{eq:BTDequantization.Dual.FinVecSpace}
			\operatorname{Tr}\bigl( A_{\hbar} T_{\hbar}( f ) \bigr)
	&= \frac{1}{( 2 \pi \hbar )^{N}}
			\int_{\mathcal{S}} \varXi_{\hbar}( A_{\hbar} ) f \,\mathrm{dvol}
	.
\end{align}
for the $2 N$-dimensional vector space $\mathcal{S}$.
Similarly to Toeplitz quantization, we may extend the domain of Berezin-Toeplitz dequantization to all bounded operators, however, the trace is only partially defined.
For all trace-class operators $A_{\hbar}$ and all complex-valued Schwartz functions $f \in \mathrm{C}^\infty_\mathrm{S}( \mathcal{S}, \mathbb{C} )$, the trace on the left hand side written in the section basis \eref{eq:HilbertSpace.Basis} becomes
\begin{align}
\label{eq:BTDequantization.Trace}
			\operatorname{Tr}\bigl( A_{\hbar} T_{\hbar}( f ) \bigr)
	&= \sum_{n_{1}, \dots, n_{N} = 0}^{\infty}
				\bra{n_{1}, \dots, n_{N}}_{\hbar}
				A_{\hbar} T_{\hbar}( f )
				\ket{n_{1}, \dots, n_{N}}_{\hbar}
	.
\end{align}
Dequantizing a projector $\ket{j_{1}, \dots, j_{N}}_{\hbar} \bra{j_{1}, \dots, j_{N}}_{\hbar}$ for any $j_{1}, \dots, j_{N} \in \mathbb{N}$ yields
\begin{align}
\label{eq:BTDequantization.BasisProjector}
				\varXi_{\hbar}\bigl(
					\ket{j_{1}, \dots, j_{N}}_{\hbar}
					\bra{j_{1}, \dots, j_{N}}_{\hbar}
				\bigr)( z )
	&= \exp\left(
			- \frac{1}{\hbar} | z |^2
			\right)
			\prod_{k = 1}^{N}
				\frac{1}{j_{k}!}
				\left( \frac{| z^{k} |^2}{\hbar} \right)^{j_{k}}
	.
\end{align}
We use this as a consistency check and notice that the identity operator $\mathbf{1} \in \mathcal{B}( \mathcal{H}_{\hbar} )$ dequantizes to the constant function $1$ when we extend the dequantization map to the domain of bounded operators and codomain of bounded functions.

We define the Berezin transform kernel $b_{\hbar}$ from the exponential factor in \eref{eq:BTDequantization.BasisProjector} such that for any $z \in \mathcal{S}$:
\begin{align}
\label{eq:VectorSpace.BTransformKernel}
			b_{\hbar}( z )
	&:= \frac{1}{( 2 \pi \hbar )^N}
			\exp\left(
			- \frac{1}{\hbar} | z |^2
			\right)
	.
\end{align}
Let $\circledast$ denote the convolution product between any pair of functions $f_{1}, f_{2} \in \mathrm{L}^1( \mathcal{S}, \mathbb{C} )$ such that
\begin{align}
\label{eq:VectorSpace.Convolution}
			( f_{1} \circledast f_{2} )( z )
	&= \int_{\mathcal{S}}
				f_{1}( z - z' )
				f_{2}( z' )
			\,\mathrm{dvol}( z' )
	.
\end{align}
By expanding Toeplitz operators in terms of the projectors $\ket{j_{1}, \dots, j_{N}}_{\hbar} \bra{j_{1}, \dots, j_{N}}_{\hbar}$, we derive the explicit expression for the Berezin transform of any Schwartz function $f \in \mathrm{C}^\infty_\mathrm{S}( \mathcal{S}, \mathbb{C} )$ (or even any bounded function) as a convolution with the Berezin kernel \eref{eq:VectorSpace.BTransformKernel},
\begin{align}
\label{eq:VectorSpace.BTransform}
			( \varXi_{\hbar} \after T_{\hbar} )( f )
	&= b_{\hbar} \circledast f
	,
\end{align}
which is also an element of $\mathrm{C}^\infty_\mathrm{b}( \mathcal{S}, \mathbb{C} )$.
This follows also from the integral kernels of the projectors, see also \cite[Eq.~4.1.84]{2007MaMarinescu} and \cite[Eq.~1.91]{2008MaMarinescu-Bergman}, since for all $f, g \in \mathrm{C}^\infty_\mathrm{S}( \mathcal{S}, \mathbb{C} )$: 
\begin{subequations}
\begin{align}
		  \operatorname{Tr}\bigl( T_{\hbar}( f ) T_{\hbar}( g ) \bigr)
	&= \int_{\mathcal{S}} ( \varXi_{\hbar} \after T_{\hbar} )( f )( z ) g( z ) \,\mathrm{d}{\mu_{\hbar}( z )}
\allowdisplaybreaks[3]\\
	&= \iint_{\mathcal{S}} \varPi_{\hbar}( z, z' ) f( z' ) \varPi_{\hbar}( z', z ) g( z )
 		  \,\mathrm{d}{\mu_{\hbar}}( z ) \,\mathrm{d}{\mu_{\hbar}}( z' )
\allowdisplaybreaks[3]\\
		  ( \varXi_{\hbar} \after T_{\hbar} )( f )( z )
	&= \int_{\mathcal{S}} \underbrace{\varPi_{\hbar}( z', z ) \varPi_{\hbar}( z, z' )}_{b_{\hbar}( z - z' )} f( z' )
 		  \,\mathrm{d}{\mu_{\hbar}}( z' )
    .
\end{align}
\end{subequations}

There is an example for a Berezin transform given in \ref{sec:AppendixNormalAntiNormalOrdering}.
Note that in the classical limit $\hbar \to 0$, the Gaussian function \eref{eq:VectorSpace.BTransformKernel} converges to a delta distribution -- the identity with respect to convolution, thus
\begin{align}
\label{eq:VectorSpace.BTransform.ClassicalLimit}
			\lim_{\hbar \to 0}
				( \varXi_{\hbar} \after T_{\hbar} )( f )
	&= f
	.
\end{align}

These observations are useful for the construction of a continuous field of C*-algebras (including the classical limit $\hbar = 0$) further below.

\section{The Sorkin-Johnston state, and the star products of (Berezin-)Toeplitz (de)quantization}
\label{sec:SJStateAndStrictDeformationQuantization}
To define a state with dequantization and prove that it is the same quasi-free state as the Sorkin-Johnston state, we first need to understand the relationship between the Weyl quantization and the Berezin-Toeplitz dequantization map. 
We also use the Weyl generators in the proofs of strict deformation quantizations further below. 

\subsection{The Weyl algebra and its relation to (Berezin-)Toeplitz (de)quantization}
\begin{lemma}
	% Let $\mathfrak{W}_{\hbar}$ be the Weyl C*-algebra over the vector space $\mathcal{S}$ with Weyl generators labelled by covectors
	For $\phi \in \mathcal{S}^* = \operatorname{Hom}( \mathcal{S}, \mathbb{R} )$, denote  the complex components as $\phi_{i} \in \mathbb{C}$ such that (in the summation convention)
	\begin{align}
	\label{eq:CovectorComponents}
				\phi( z )
		&= \phi_{i} z^{i}
			+ \conj\phi_{\ib} \conj{z}^{\ib}
		.
	\end{align}
	With this notation, denote a corresponding linear combination of the creation $a^{+}_{\jb}$ and annihilation operators $a^{-}_{j}$ by
	\begin{align}
	\label{eq:WeylGenerator.Exponent}
				\varPhi_{\hbar}( \phi )
		&:= \sqrt{\hbar} \;
				\delta^{i \ib}
				\left(
					\phi_{i} a^{+}_{\ib}
				+ \conj\phi_{\ib} a^{-}_{i}
				\right)
		.
	\end{align}
	Define a function $W_{\hbar} : \mathcal{S}^* \to \mathcal{B}( \mathcal{H}_{\hbar} )$ by
	\begin{align}
	\label{eq:WeylGenerator}
				W_{\hbar}( \phi )
		&:= \exp\bigl(
					\mathrm{i}
					\varPhi_{\hbar}( \phi )
				\bigr)
	    .
	\end{align}
	These operators $W_{\hbar}( \phi )$ fulfill the Weyl relations given in \eref{eq:WeylRelations}.
\end{lemma}
\proof
The unit of the Weyl algebra \eref{eq:WeylRelations.Unit} is obviously given by $W_{\hbar}( 0 )$.
For the involution \eref{eq:WeylRelations.Involution}, note that $( a^{\pm}_{j} )^* = a^{\mp}_{j}$ and thus \eref{eq:WeylGenerator.Exponent} is self-adjoint.
Hence
\begin{align}
\label{eq:WeylGenerator.Adjoint}
            W_{\hbar}( \phi )^*
    &= W_{\hbar}( -\phi )
    .
\end{align}
For the product of two generators \eref{eq:WeylRelations.Product}, compute the commutator
\begin{subequations}
\begin{align}
\label{eq:WeylGenerator.ExponentCommutator}
            \comm{\varPhi_{\hbar}( \phi )}{\varPhi_{\hbar}( \phi' )}
    &= \hbar \delta^{i \ib} \delta^{j \jb}
            \comm{
                \phi_{i} a^{+}_{\ib} + \conj\phi_{\ib} a^{-}_{i}
            }{
                \phi'_{j} a^{+}_{\jb} + \conj\phi'_{\jb} a^{-}_{j}
            }
    \allowdisplaybreaks[3]\\
    &= - \hbar \delta^{i \ib} \delta^{j \jb}
            \left(
                \phi_{i} \conj\phi'_{\jb}
            - \conj\phi_{\ib} \phi'_{j}
            \right)
            \delta_{\ib j}
            \mathbf{1}
    \allowdisplaybreaks[3]\\\label{eq:WeylGenerator.ExponentPoissonBracket}
    &= \mathrm{i} \hbar \PoiBr{\phi}{\phi'} \mathbf{1}
    .
\end{align}
\end{subequations}
It is seen that the commutator of this expression with $\varPhi_{\hbar}( \phi )$ vanishes so that the Baker-Campbell-Hausdorff formula yields
\begin{align}
\label{eq:BakerCampbellHausdorff}
            \exp\bigl(
                \mathrm{i}
                \varPhi_{\hbar}( \phi )
            \bigr)
            \exp\bigl(
                \mathrm{i}
                \varPhi_{\hbar}( \phi' )
            \bigr)
    &= \exp\left( -\frac{1}{2} \comm{\varPhi_{\hbar}( \phi )}{\varPhi_{\hbar}( \phi' )} \right)
            \exp\bigl(
                \mathrm{i}
                \varPhi_{\hbar}( \phi + \phi' )
            \bigr)
    .
\end{align}
Replace the commutator in \eref{eq:BakerCampbellHausdorff} with expression \eref{eq:WeylGenerator.ExponentPoissonBracket} to show that the generators \eref{eq:WeylGenerator} also fulfill \eref{eq:WeylRelations.Product}, and thus all Weyl relations \eref{eq:WeylRelations}.
\qed

The Berezin-Toeplitz quantization respects anti-normal ordering and dequantization respects normal ordering (see also \ref{sec:AppendixNormalAntiNormalOrdering}).
To reorder the terms in the series expansion of a Weyl generator \eref{eq:WeylGenerator}, we use the commutation relations \eref{eq:HilbertSpace.CreationAnnihilationCommutators} and derive commutators for powers of ladder operators in anti-normal or normal order.
For an index pair $( i, \ib ) \in \{ ( 1, \bar{1} ), ( 2, \bar{2} ), \dotso, ( N, \bar{N} ) \}$, the two orders of the commutators are, respectively,
\begin{subequations}
\label{eq:HilbertSpace.CreationAnnihilationHigherCommutators}
\begin{align}
\label{eq:HilbertSpace.CreationAnnihilationHigherCommutatorsAntiNormal}
			\comm{( a^{-}_{i} )^{m}}{( a^{+}_{\ib} )^{n}}
	&= \sum_{l = 1}^{\min( m, n )}
			(-1)^{l + 1}
			l! \binom{m}{l} \binom{n}{l}
			( a^{-}_{i} )^{m - l}
			( a^{+}_{\ib} )^{n - l}
	,
	\allowdisplaybreaks[3]\\
\label{eq:HilbertSpace.CreationAnnihilationHigherCommutatorsNormal}
			\comm{( a^{-}_{i} )^{m}}{( a^{+}_{\ib} )^{n}}
	&= \sum_{l = 1}^{\min( m, n )}
				l! \binom{m}{l} \binom{n}{l}
				( a^{+}_{\ib})^{n - l}
				( a^{-}_{i})^{m - l}
	,
\end{align}
\end{subequations}
while for all index pairs $( i, \ib ) \notin \{ ( 1, \bar{1} ), ( 2, \bar{2} ), \dotso, ( N, \bar{N} ) \}$ these commutators vanish.
For the second order term, for example, the reorder yields one extra term that is the same for quantization and dequantization but with opposite sign,
\begin{subequations}
\label{eq:WeylGenerator.SecondOrderTerm}
\begin{align}
			\bigl( \mathrm{i} \varPhi_{\hbar}( \phi ) \bigr)^2
	&= - \hbar
			\delta^{i \ib}
			\delta^{j \jb}
			\bigl(
				\underbrace{\phi_{i} \phi_{j} a^{+}_{\ib} a^{+}_{\jb}
				+ 2 \conj\phi_{\ib} \phi_{j} a^{-}_{i} a^{+}_{\jb}
				+ \conj\phi_{\ib} \conj\phi_{\jb} a^{-}_{i} a^{-}_{j}
				}_{\text{anti-normal ordered, by $T$-quantization}}{}
			- \conj\phi_{\ib} \phi_{j} \delta_{i \jb} \mathbf{1}
			\bigr)
	\allowdisplaybreaks[3]\\
	&= - \hbar
			\delta^{i \ib}
			\delta^{j \jb}
			\bigl(
				\underbrace{\phi_{i} \phi_{j} a^{+}_{\ib} a^{+}_{\jb}
				+ 2 \phi_{i} \conj\phi_{\jb} a^{+}_{\ib} a^{-}_{j}
				+ \conj\phi_{\ib} \conj\phi_{\jb} a^{-}_{i} a^{-}_{j}
				}_{\text{normal ordered, for $\varXi$-dequantization}}{}
			+ \phi_{i} \conj\phi_{\jb} \delta_{i \jb} \mathbf{1}
			\bigr)
	.
\end{align}
\end{subequations}
The extra terms of all orders yield an exponential amplitude factor depending on $\hbar$ and $| \phi |^2 = \sum_{i = 1}^{N} | \phi_{i} |^2$,
\begin{align}
\label{eq:WeylGenerator.BTQuantization}
			W_{\hbar}( \phi )
	&= \exp\left(
				\frac{\hbar}{2} | \phi |^2
			\right)
			T_{\hbar}\left(
				\mathrm{e}^{\mathrm{i} \phi}
			\right)
	.
\end{align}
Similarly, dequantization of the Weyl generators gives
\begin{align}
\label{eq:WeylGenerator.BTDequantization}
			\varXi_{\hbar}\bigl( W_{\hbar}( \phi ) \bigr)
	&= \exp\left(
			- \frac{\hbar}{2} | \phi |^2
			\right)
			\mathrm{e}^{\mathrm{i} \phi}
	.
\end{align}
Like the Toeplitz sections given in \eref{eq:Quantization.Section}, every Weyl generator forms a vector field in $\prod_{\hbar \in I} \mathcal{A}_{\hbar}$.

\begin{definition}
	The \defof{Weyl section} of a covector $\phi \in \mathcal{S}^* = \operatorname{Hom}( \mathcal{S}, \mathbb{R} )$ is
	\begin{align}
	\label{eq:WeylSection}
				W( \phi ) : \hbar
		&\mapsto \begin{cases}
					\mathrm{e}^{\mathrm{i} \phi} & \hbar = 0
				, \\
					W_{\hbar}( \phi ) & \hbar > 0 .
				\end{cases}
	\end{align}
	% over the family of C*-algebras of bounded operators $\mathcal{B}( \mathcal{H}_{\hbar} )$ for $\hbar \in \mathbb{R}_{*}$, and bounded functions $\mathrm{C}_\mathrm{b}( \mathcal{S}, \mathbb{C} )$ at $\hbar = 0$.
\end{definition}
In the limit $\hbar \to 0$, the quantization \eref{eq:WeylGenerator.BTQuantization} and dequantization \eref{eq:WeylGenerator.BTDequantization} coincide with $\exp( \mathrm{i} \phi )$, which is used in showing that the Weyl map is a strict deformation quantization and also continuous in the classical limit.

Any Toeplitz operator of a Schwartz function $f \in \mathrm{C}^\infty_\mathrm{S}( \mathcal{S}, \mathbb{C} )$ can also be written in terms of Weyl generators.
For this, consider the Fourier transform, which is an automorphism on $\mathrm{C}^\infty_\mathrm{S}( \mathcal{S}, \mathbb{C} )$,
\begin{align}
\label{eq:VectorSpace.FourierTransform}
			\hat{f}( \phi )
	&= \frac{1}{( 2 \pi )^{2 N}}
			\int_{\mathcal{S}}
				f( z )
				\mathrm{e}^{-\mathrm{i} \phi( z )}
			\,\mathrm{dvol}( z )
	,
\end{align}
and its inverse transform (with the volume form $\mathrm{dvol}^*$ on $\mathcal{S}^*$),
\begin{align}
\label{eq:VectorSpace.FourierTransformInverse}
			f( z )
	&= \int_{\mathcal{S}^*}
				\hat{f}( \phi )
				\mathrm{e}^{\mathrm{i} \phi( z )}
			\,\mathrm{dvol}^*( \phi )
	.
\end{align}
Recall that the result \eref{eq:WeylGenerator.BTDequantization} is related to \eref{eq:WeylGenerator.BTQuantization} by a Berezin transform.
The Weyl quantization is related to ``half'' a Berezin transform, so consider
\begin{align}
\label{eq:VectorSpace.BRootFourierTransform}
			\sqrt{\widehat{b_{\hbar}}}( \phi )
	&= \frac{1}{( 2 \pi )^{N}}
			\exp\left(
			- \frac{\hbar}{2} | \phi |^2
			\right)
	.
\end{align}
Note that the exponential function here is exactly the same as in the dequantization \eref{eq:WeylGenerator.BTDequantization}.
The Toeplitz operator of $f$ is then
\begin{subequations}
\label{eq:BTQuantization.WeylGenerators}
\begin{align}
			T_{\hbar}( f )
	&= \int_{\mathcal{S}^*}
				\hat{f}( \phi )
				T_{\hbar}\left( \mathrm{e}^{\mathrm{i} \phi} \right)
			\,\mathrm{dvol}^*( \phi )
	\\
	&= ( 2 \pi )^{N}
			\int_{\mathcal{S}^*}
				\hat{f}( \phi )
				\sqrt{\widehat{b_{\hbar}}}( \phi )
				W_{\hbar}( \phi )
			\,\mathrm{dvol}^*( \phi )
	.
\end{align}
\end{subequations}
Note that in the classical limit $\hbar \to 0$, $f_{\hbar} \to f$ similarly to the limit of the Berezin transform, the left hand side of \eref{eq:BTQuantization.WeylGenerators} becomes $f$, and the right hand side of \eref{eq:BTQuantization.WeylGenerators} becomes the inverse Fourier transform \eref{eq:VectorSpace.FourierTransformInverse}.
More details on the Weyl algebra and its relation to Toeplitz operators are given in \cite[ch.~II]{1998Landsman}.

\subsection{The Sorkin-Johnston state from dequantization}
\label{subsec:DequantizationState}
Now we use dequantization to define a state and compare its properties to the Sorkin-Johnston state using the Weyl algebra and the relation to Berezin-Toeplitz dequantization.
\begin{theorem}
	The linear map $\sigma_{\hbar} : \mathfrak{A}_{\hbar} \to \mathbb{C}$ given by
	\begin{align}
	\label{eq:BTDequantization.State}
				\sigma_{\hbar}( A )
		&:= \varXi_{\hbar}( A )( 0 )
	\end{align}
	is the Sorkin-Johnston state.
\end{theorem}
\proof
In order to show that this map is the Sorkin-Johnston state \eref{eq:SJState}, we need to evaluate it on Weyl generators.
Recall the result \eref{eq:WeylGenerator.BTDequantization} when dequantizing the Weyl generator $W_{\hbar}( \phi )$ of any covector $\phi \in \mathcal{S}^* = \operatorname{Hom}( \mathcal{S}, \mathbb{R} )$.
Evaluation at $0 \in \mathcal{S}$ yields
\begin{align}
\label{eq:SJState.WeylGenerator}
            \sigma_{\hbar}\bigl( W_{\hbar}( \phi ) \bigr)
    &= \exp\left(
            - \frac{\hbar}{2} | \phi |^2
            \right)
    .
\end{align}
In order to compare it with \eref{eq:SJState}, notice that
\begin{align}
    \label{eq:SJState.CovarianceCalculation}
           \eta^{-1}( \phi, \phi )
    &= \delta^{i \ib} \phi_{i} \conj\phi_{\ib}
        + \delta^{\ib i} \conj\phi_{\ib} \phi_{i}
    = 2 | \phi |^2
\end{align}
so that
\begin{align}
\label{eq:SJState.Covariance}
            | \phi |^2
    &= \frac{1}{2} \eta^{-1}( \phi, \phi )
    .
\end{align}
We obtain the inverse of the bi-linear form $\eta$ on $\mathcal{S}$, which is identical to the covariance of the Sorkin-Johnston state \eref{eq:SJState}.
The form $\eta$ is compatible with the complex structure $J$ yielding a K\"ahler vector space $\smash( \mathcal{S}, \omega, \eta, J )$.
\qed

For any Toeplitz operator $T_{\hbar}( f ) \in \mathfrak{A}_{\hbar}$, the Sorkin-Johnston state $\sigma_{\hbar}$ is the Berezin transform of $f \in \mathcal{A}_{0}$ evaluated at 0,
\begin{align}
\label{eq:BTDequantization.StateBerezinTransform}
			\sigma_{\hbar}\bigl( T_{\hbar}( f ) \bigr)
	&= \int_{\mathcal{S}}
				b_{\hbar}( z ) f( z )
			\,\mathrm{dvol}( z )
	.
\end{align}

Note that the dequantization state is parametrized by $\hbar$, so $( \sigma_{\hbar} )_{\hbar \in I}$ is a family of states with the classical limit $\sigma_{0}( f ) := f( 0 )$.
For any section $A$ of the continuous field of C*-algebras $\bigl( I, ( \mathfrak{A}_{\hbar} )_{\hbar \in I}, \varGamma \bigr)$, the map $\sigma( A ) : I \to \mathbb{C}$ given by
\begin{align}
\label{eq:BTDequantization.StateField}
			\sigma( A )( \hbar )
	&:= \sigma_{\hbar}\bigl( A( \hbar ) \bigr)
\end{align}
is continuous since $\varXi( A ) : \mathcal{M} \times I \to \mathbb{C}$ defined as in \eref{eq:Dequantization.Field} is continuous.

\subsection{Infinite order strict deformation (de)quantization}
\label{subsec:InfiniteOrderStrictDeformations}
Let $( \mathfrak{A}_{\hbar} )_{\hbar \in I}$ be the family of C*-algebras with $\mathfrak{A}_{0} = \mathrm{C}_0( \mathcal{S}, \mathbb{C} )$ --- as the closure of $\mathcal{A}_{0} = \mathrm{C}^\infty_\mathrm{S}( \mathcal{S}, \mathbb{C} )$ --- and compact operators $\mathfrak{A}_{\hbar} = \mathcal{K}( \mathcal{H}_{\hbar} )$ for $\hbar \in I_{*}$. There exists a continuous field of C*-algebras $\bigl( I, ( \mathfrak{A}_{\hbar} )_{\hbar \in I}, \varGamma \bigr)$ equivalently determined by the Weyl sections and the Toeplitz sections \cite[Ch.~II, Sec.~2.6]{1998Landsman}.
As an instance of our general discussion of dequantization-expandibility in \Defref{def:CField.Expandability}, we will show that Toeplitz sections of Schwartz functions are $\varXi$-dequantization expandable sections.

In the proofs below, we have to bound $k$-th order remainders $\mathrm{er}_{k} \in \mathrm{C}^\infty( \mathbb{C}, \mathbb{C} )$ for the Taylor expansion of the exponential function,
\begin{align}
\label{eq:ExponentialRemainder}
			\mathrm{er}_{k}( \zeta )
	&= \mathrm{e}^{\zeta}
		- \sum_{i = 0}^{k}
				\frac{\zeta^{j}}{j!}
	.
\end{align}

\begin{lemma}
\label{lma:ExponentialRemainder.Bound}
	For every $k \in \mathbb{N}$, there exists a real constant $C_{k} > 0$ such that for all $\zeta \in \mathbb{C}$
	\begin{align}
	\label{eq:ExponentialRemainder.Bound}
				\bigl| \mathrm{er}_{k}( \zeta ) \bigr|
		&\leq C_{k}
				\left( 1 + \mathrm{e}^{\operatorname{Re} \zeta} \right)
				| \zeta |^{k + 1}
		.
	\end{align}
\end{lemma}
\proof
Taylor's theorem states that
\begin{align}
\label{eq:ExponentialRemainder.TaylorBound}
            \bigl| \mathrm{er}_{k}( \zeta ) \bigr|
    &= \mathcal{O}\left( | \zeta |^{k + 1} \right)
\end{align}
as $| \zeta |$ becomes small.
To find a bound for $| \zeta | \to \infty$, use the triangle inequality,
\begin{align}
\label{eq:ExponentialRemainder.TriangularBound}
            \bigl| \mathrm{er}_{k}( \zeta ) \bigr|
    &\leq \mathrm{e}^{\operatorname{Re} \zeta}
        + \sum_{j = 0}^{k} \frac{| \zeta |^{k}}{j!}
    .
\end{align}
We add the two bounds \eref{eq:ExponentialRemainder.TaylorBound} and \eref{eq:ExponentialRemainder.TriangularBound} together to obtain \eref{eq:ExponentialRemainder.Bound}.
\qed

\begin{proposition}
\label{prop:BTSection.DequantizationExpandable}
	Given any Schwartz function $f \in \mathrm{C}^\infty_\mathrm{S}( \mathcal{S}, \mathbb{C} )$, the corresponding Toeplitz section $T( f )$ is $\varXi$-expandable.
\end{proposition}
\proof
Use the Toeplitz section $A = T( f )$ in \eref{eq:CField.DequantizationExpandable} --- setting the dequantization $\varUpsilon = \varXi$ --- to obtain a condition for the Berezin transform of $f$,
\begin{align}
\label{eq:BTSection.DequantizationExpandable}
            \lim_{\hbar \to 0}
                \frac{1}{\hbar^{k}}
                \left\|
                    ( \varXi_{\hbar} \after T_{\hbar} )( f )
                - \sum_{j = 0}^{k}
                        f_{j}
                        \hbar^{j}
                \right\|
    &= 0
    .
\end{align}
This condition is fulfilled by the functions
\begin{align}
\label{eq:BTSection.DequantizationExpandableCoeff}
            f_{j}( z )
    &= \frac{1}{j!}
            \left(
                \delta^{i \ib}
                \pderiv{}{z^{i}} \pderiv{}{\conj{z}^{i}}
            \right)^{j}
            f( z )
\end{align}
for all orders $j, k \in \mathbb{N}$.
To show that the functions $f_{j}$ indeed fulfill the condition, first consider a Schwartz function $f \in \mathrm{C}^\infty_\mathrm{S}( \mathcal{S}, \mathbb{C} )$, use the Fourier transforms $\hat{f}$ and $\hat{b}_{\hbar}( \phi ) = ( 2 \pi )^{2 N} \exp( -\hbar | \phi |^2 )$ with the convolution theorem.
The derivatives in \eref{eq:BTSection.DequantizationExpandableCoeff} become $\mathrm{i} \phi_{i}$ and $\mathrm{i} \conj\phi_{\ib}$, respectively, so
\begin{subequations}
\label{eq:BTSection.DequantizationExpandableNorm}
\begin{align}
            \left\|
                ( \varXi_{\hbar} \after T_{\hbar} )( f )
            - \sum_{j = 0}^{k}
                    f_{j}
                    \hbar^{j}
            \right\|
    &= \left\| \;
                \int_{\mathcal{S}^*}
                    \mathrm{er}_{k}\left(
                    - \hbar | \phi |^2
                    \right)
                    \hat{f}( \phi )
                    \mathrm{e}^{\mathrm{i} \phi}
                \,\mathrm{dvol}^*( \phi )
            \right\|
    \\
    &\leq \int_{\mathcal{S}^*}
                \left|
                    \mathrm{er}_{k}\left(
                    - \hbar | \phi |^2
                    \right)
                \right|
                \left| \hat{f}( \phi ) \right|
            \,\mathrm{dvol}^*( \phi )
    .
\end{align}
\end{subequations}
Now apply \Lmaref{lma:ExponentialRemainder.Bound} and note that here the argument $\zeta$ is non-positive, such that
\begin{align}
\label{eq:BTSection.DequantizationExpandableNorm2}
            \left\|
                ( \varXi_{\hbar} \after T_{\hbar} )( f )
            - \sum_{j = 0}^{k}
                    f_{j}
                    \hbar^{j}
            \right\|
    &\leq 2 C_{k} \hbar^{k + 1}
            \int_{\mathcal{S}^*}
                | \phi |^{2 ( k + 1 )}
                \left| \hat{f}( \phi ) \right|
            \,\mathrm{dvol}^*( \phi )
    .
\end{align}
When $f$ is Schwartz, then $\hat{f}$ is Schwartz, the integral is finite, and we obtain an upper bound given by some finite constant times $\hbar^{k + 1}$.
So the limit expression \eref{eq:BTSection.DequantizationExpandable} vanishes for all $k \in \mathbb{N}$.
\qed

The Toeplitz quantization star product $\star_{T}$ for Schwartz functions $f_1, f_2 \in \mathrm{C}^\infty_\mathrm{S}( \mathcal{S}, \mathbb{C} )$ is determined by the conditions \eref{eq:Quantization.StarProductCondition}.
It has an exponential expression with directed derivatives that act only on the function to the left or right as indicated with an arrow,
\begin{align}
\label{eq:BTQuantization.StarProduct}
			f_1 \star_{T} f_2
	&= f_1
			\exp\left(
			- \hbar
				\overset{\leftarrow}{\pderiv{}{z^{i}}}
				\delta^{i \ib}
				\overset{\rightarrow}{\pderiv{}{\conj{z}^{\ib}}}
			\right)
			f_2
	.
\end{align}
Similarly, the dequantization star product $\star_{\varXi}$ determined by the conditions \eref{eq:Dequantization.StarProductCondition} is a star product for functions $f_1, f_2 \in \mathrm{C}^\infty_\mathrm{S}( \mathcal{S}, \mathbb{C} )$ with the exponential expression
\begin{align}
\label{eq:BTDequantization.StarProduct}
			f_1 \star_{\varXi} f_2
	&= f_1
			\exp\left(
				\hbar
				\overset{\leftarrow}{\pderiv{}{\conj{z}^{\ib}}}
				\delta^{\ib i}
				\overset{\rightarrow}{\pderiv{}{z^{i}}}
			\right)
			f_2
	.
\end{align}
Note that the holomorphic and anti-holomorphic derivatives act in different directions and the exponentials have opposite sign.

\begin{proposition}
\label{prop:BTQuantization.StrictDeformation}
	Toeplitz quantization with the star product \eref{eq:BTQuantization.StarProduct} is an infinite order strict deformation quantization over the algebra of Schwartz functions $\mathcal{A}_{0} = \mathrm{C}^\infty_\mathrm{S}( \mathcal{S}, \mathbb{C} )$.
\end{proposition}
\proof
In \cite[Ch.~II, Sec.~2.6]{1998Landsman}, it was shown that there exists a continuous field of C*-algebras $\bigl( I, ( \mathfrak{A}_{\hbar} )_{\hbar \in I}, \varGamma \bigr)$ including the Toeplitz sections $T( f )$ for $f \in \mathfrak{A}_{0} = \mathrm{C}_0( \mathcal{S}, \mathbb{C} )$ as sections, $T( f ) \in \varGamma$.
It remains to show that the star product \eref{eq:BTQuantization.StarProduct} fulfills the conditions \eref{eq:Quantization.StarProductCondition} in all orders $k \in \mathbb{N}$.

Recall the Fourier decomposition \eref{eq:BTQuantization.WeylGenerators} of the Toeplitz operators $T_{\hbar}( f )$ and $T_{\hbar}( f' )$ for any functions $f, f' \in \mathrm{C}^\infty_\mathrm{S}( \mathcal{S}, \mathbb{C} )$ into Toeplitz operators $T_{\hbar}( \mathrm{e}^{\mathrm{i} \phi} )$ and $T_{\hbar}( \mathrm{e}^{\mathrm{i} \phi'} )$ (or Weyl generators).
The $k$-th remainder \eref{eq:Quantization.StarProductRemainder} is then bounded from above by the double integral
\begin{align}
\label{eq:BTQuantization.StarProductRemainder}
                R_{T}^{k}( f, f', \hbar )
    &\leq \iint_{\mathcal{S}^*}
                \left| \hat{f}( \phi ) \right|
                \left| \hat{f}'( \phi' ) \right|
                R_{T}^{k}\left(
                    \mathrm{e}^{\mathrm{i} \phi}, \mathrm{e}^{\mathrm{i} \phi'}, \hbar
                \right)
            \,\mathrm{dvol}^*( \phi )
            \,\mathrm{dvol}^*( \phi' )
    .
\end{align}
The norm inside the integral is given by
\begin{align}
\label{eq:BTQuantization.StarProductFourierRemainder}
            R_{T}^{k}\left(
                \mathrm{e}^{\mathrm{i} \phi}, \mathrm{e}^{\mathrm{i} \phi'}, \hbar
            \right)
    &= \frac{1}{\hbar^k}
            \left\|
                T_{\hbar}\left( \mathrm{e}^{\mathrm{i} \phi} \right)
                T_{\hbar}\left( \mathrm{e}^{\mathrm{i} \phi'} \right)
            - \sum_{j = 0}^{k}
                    \frac{\hbar^{j}}{j!}
                    \left( \delta^{i \ib} \phi_{i} \conj\phi'_{\ib} \right)^{j}
                    T_{\hbar}\left( \mathrm{e}^{\mathrm{i} ( \phi + \phi' )} \right)
            \right\|
    ,
\end{align}
where the Weyl relations imply
\begin{align}
\label{eq:BTQuantization.BakerCampbellHausdorff}
            T_{\hbar}\left( \mathrm{e}^{\mathrm{i} \phi} \right)
            T_{\hbar}\left( \mathrm{e}^{\mathrm{i} \phi'} \right)
    &= \exp\left( \hbar \delta^{i \ib} \phi_{i} \conj\phi'_{\ib} \right)
            T_{\hbar}\left( \mathrm{e}^{\mathrm{i} ( \phi + \phi' )} \right)
    .
\end{align}
To apply \Lmaref{lma:ExponentialRemainder.Bound}, set $\zeta \in \mathbb{C}$ as
\begin{subequations}
\label{eq:BTQuantization.StarProductFourierRemainderVariables}
\begin{align}
            \zeta
    &= \hbar \delta^{i \ib}
            \phi_{i} \conj\phi'_{\ib}
    , \\
            \operatorname{Re} \zeta
    &= \frac{\hbar}{2} \delta^{i \ib}
            \left(
                \phi_{i} \conj\phi'_{\ib}
            + \conj\phi_{\ib} \phi'_{i}
            \right)
    .
\end{align}
\end{subequations}
Thus, we have
\begin{subequations}
\label{eq:BTQuantization.StarProductFourierRemainderSimplified}
\begin{align}
            R_{T}^{k}\left(
                \mathrm{e}^{\mathrm{i} \phi}, \mathrm{e}^{\mathrm{i} \phi'}, \hbar
            \right)
    &= \frac{1}{\hbar^{k}}
            \left| \mathrm{er}_{k}( \zeta ) \right|
            \left\|
                T_{\hbar}\left( \mathrm{e}^{\mathrm{i} ( \phi + \phi' )} \right)
            \right\|
    \\
    &\leq C_{k}
            \hbar
            \left| \delta^{i \ib} \phi_{i} \conj\phi'_{\ib} \right|^{k + 1}
            \Bigl(
                \left\|
                    T_{\hbar}\left( \mathrm{e}^{\mathrm{i} ( \phi + \phi' )} \right)
                \right\|
            + \mathrm{e}^{\operatorname{Re} \zeta}
                \left\|
                    T_{\hbar}\left( \mathrm{e}^{\mathrm{i} ( \phi + \phi' )} \right)
                \right\|
            \Bigr)
    .
\end{align}
\end{subequations}
The two terms with the operator norm follow from \eref{eq:WeylGenerator.BTQuantization} and $\| W_{\hbar}( \phi ) \| = 1$ (for any $\phi \in \mathcal{S}^*$), implying that the Toeplitz map is norm contracting,
\begin{subequations}
\label{eq:BTQuantization.ToeplitzNorm}
\begin{align}
\label{eq:BTQuantization.ToeplitzNormGaussian}
            \left\| T_{\hbar}\left( \mathrm{e}^{\mathrm{i} ( \phi + \phi' )} \right) \right\|
    &= \exp\left(
            - \frac{\hbar}{2} | \phi + \phi' |^2
            \right)
    ,
    \\\label{eq:BTQuantization.ToeplitzNormDoubleGaussian}
            \mathrm{e}^{\operatorname{Re} \zeta}
            \left\| T_{\hbar}\left( \mathrm{e}^{\mathrm{i} ( \phi + \phi' )} \right) \right\|
    &= \exp\left(
            - \frac{\hbar}{2} | \phi |^2
            - \frac{\hbar}{2} | \phi' |^2
            \right)
    .
\end{align}
\end{subequations}
Both of these exponentials are bounded by 1.
So \eref{eq:BTQuantization.StarProductFourierRemainderSimplified} is bounded by
\begin{subequations}
\label{eq:BTQuantization.StarProductFourierRemainderBounded}
\begin{align}
            R_{T}^{k}\left(
                \mathrm{e}^{\mathrm{i} \phi}, \mathrm{e}^{\mathrm{i} \phi'}, \hbar
            \right)
    &\leq 2 C_{k}
            \hbar
            \left| \delta^{i \ib} \phi_{i} \conj\phi'_{\ib} \right|^{k + 1}
    .
\end{align}
\end{subequations}
The modulus in the $( k + 1 )$ order polynomial is bounded from above by the sum of $| \phi_{i} |$ and $| \phi'_{i} |$ all to the power of $k + 1$.
Inserted back into the integration \eref{eq:BTQuantization.StarProductRemainder} yields
\begin{align}
\label{eq:BTQuantization.StarProductRemainderBounded}
                R_{T}^{k}( f, f', \hbar )
    &\leq 2 C_{k} \hbar
            \iint_{\mathcal{S}^*}
                \left( \sum_{i = 1}^{N}
                    \left| \phi_{i} \right|
                    \left| \phi'_{i} \right|
                \right)^{k + 1}
                \left| \hat{f}( \phi ) \right|
                \left| \hat{f}'( \phi' ) \right|
            \,\mathrm{dvol}^*( \phi )
            \,\mathrm{dvol}^*( \phi' )
    .
\end{align}
The factor with the sum is a polynomial in $\phi$ and $\phi'$ and the integration with the Schwartz functions $\hat{f}$ and $\hat{f}'$ is finite.
Therefore, the remainder $R_{T}^{k}( f, f', \hbar )$ is bounded by a constant (independent of $\hbar$) times $\hbar$, which vanishes in the limit $\hbar \to 0$ for all $k \in \mathbb{N}$.

Poisson compatibility of this star product follows from the first order terms
\begin{subequations}
\begin{align}
\label{eq:BTQuantization.StarProductFirstOrder}
            f \star_{T} f' - f' \star_{T} f
    &= - \hbar
            \sum_{i = 1}^{N}
            \left(
                \pderiv{f}{z^{i}} \pderiv{f'}{\conj{z}^{i}}
            - \pderiv{f'}{z^{i}} \pderiv{f}{\conj{z}^{i}}
            \right)
        + \mathcal{O}\left( \hbar^2 \right)
    \\\label{eq:BTQuantization.StarProductPoissonCompatibility}
    &= \mathrm{i} \hbar
            \PoiBr{f}{f'}
        + \mathcal{O}\left( \hbar^2 \right)
    .
\end{align}
\end{subequations}
Notice that the star product is also self-adjoint and differential, which follows immediately from the differential form \eref{eq:BTQuantization.StarProduct}.
\qed

\begin{proposition}
\label{prop:BTDequantization.StrictDeformation}
	Berezin-Toeplitz dequantization with the star product \eref{eq:BTDequantization.StarProduct} is a strict deformation dequantization of the algebra of Schwartz functions $\mathcal{A}_{0} = \mathrm{C}^\infty_\mathrm{S}( \mathcal{S}, \mathbb{C} )$.
\end{proposition}
\proof
According to \cite[Ch.~II, Thm.~2.6.5]{1998Landsman}, the continuous fields of the Weyl quantization and Berezin-Töplitz quantization coincide.

Even though the Weyl operators $W_{\hbar}( \phi )$ are not elements of $\mathfrak{A}_{\hbar} = \mathcal{K}( \mathcal{H}_{\hbar} )$, they are $\varXi$-expandable following a similar argument as in \Propref{prop:BTSection.DequantizationExpandable} with the coefficients
\begin{align}
\label{eq:WeylSection.DequantizationExpandableCoeff}
            w^{\phi}_{j}
    &= \frac{1}{j!}
            \left(
            - \frac{1}{2} | \phi |^2
            \right)^{j}
            \mathrm{e}^{\mathrm{i} \phi}
    .
\end{align}
So, we write a $\varXi$-expandable section $A_{f} \in \varGamma$ as
\begin{align}
\label{eq:CField.BTDequantizationSection}
            A_{f}( \hbar )
    &= \int_{\mathcal{S}^*}
                \hat{f}( \phi, \hbar )
                W_{\hbar}( \phi )
            \,\mathrm{dvol}^*( \phi )
\end{align}
with a continuous amplitude function $\hat{f} \in \mathrm{C}( \mathcal{S}^* \times I, \mathbb{C} )$ such that $f( \cdot, \hbar ) \in \mathcal{A}_{0}$ for all $\hbar \in I$.
The conditions of the continuous field of C*-algebras $\bigl( I, ( \mathfrak{A}_{\hbar} )_{\hbar \in I}, \varGamma \bigr)$ in \Defref{def:CField} imply that all sections of the form $A_{f}$ span a total subspace of $\varGamma$.
This means that for any section $A \in \varGamma$ there exists such a $\varXi$-expandable section $A_{f} \in \varGamma$ such that for all $\delta > 0$ there exists a neighborhood $N_{0} \subset \mathbb{R}_{+}$ around $\hbar = 0$ such that for all $\hbar' \in N_{0}: \| A( \hbar' ) - A_{f}( \hbar' ) \| \leq \delta$.

Taking the dequantization yields the smooth function
\begin{align}
\label{eq:CField.BTDequantizationSection.Dequantization}
            \varXi_{\hbar}\bigl( A_{f}( \hbar ) \bigr)
    &= \int_{\mathcal{S}^*}
                \hat{f}( \phi, \hbar )
                \exp\left(
                - \frac{\hbar}{2} | \phi |^2
                \right)
                \mathrm{e}^{\mathrm{i} \phi}
            \,\mathrm{dvol}( \phi )
    ,
\end{align}
which is the convolution of the pointwise Fourier transformed function $f( \cdot, \hbar )$ with ``half'' the Berezin kernel.

Now consider the dequantization of a product $A_{f} A_{f'}$ of two such sections.
With the same identification of $\zeta$ as in \eref{eq:BTQuantization.StarProductFourierRemainderVariables}, rewrite the Weyl relations in terms of the complex conjugated value $\conj{\zeta} = \hbar \delta^{i \ib} \conj\phi_{\ib} \phi'_{i}$,
\begin{align}
\label{eq:BakerCampbellHausdorff.ForWeylDequantization}
			W_{\hbar}( \phi ) W_{\hbar}( \phi' )
	&= \mathrm{e}^{-\mathrm{i} \operatorname{Im} \conj{\zeta}}
			W_{\hbar}( \phi + \phi' )
	.
\end{align}
Similar to the previous proof, notice that
\begin{subequations}
\label{eq:BTDequantization.WeylDequantization}
\begin{align}
\label{eq:BTDequantization.WeylDequantizationGaussianWave}
            \varXi_{\hbar}\bigl(
                W_{\hbar}( \phi + \phi' )
            \bigr)
    &= \exp\left(
            - \frac{\hbar}{2} | \phi + \phi' |^2
            \right)
            \mathrm{e}^{\mathrm{i} ( \phi + \phi' )}
    ,
    \allowdisplaybreaks[3]\\\label{eq:BTDequantization.WeylDequantizationDoubleGaussianWave}
            \mathrm{e}^{\operatorname{Re} \conj{\zeta}}
            \varXi_{\hbar}\bigl(
                W_{\hbar}( \phi + \phi' )
            \bigr)
    &= \exp\left(
            - \frac{\hbar}{2} | \phi |^2
            - \frac{\hbar}{2} | \phi' |^2
            \right)
            \mathrm{e}^{\mathrm{i} ( \phi + \phi' )}
    \allowdisplaybreaks[3]\\\label{eq:BTDequantization.WeylDequantizationRealPart}
    &= \varXi_{\hbar}\bigl(
                W_{\hbar}( \phi )
            \bigr)
            \varXi_{\hbar}\bigl(
                W_{\hbar}( \phi' )
            \bigr)
    .
\end{align}
\end{subequations}
We combine the exponentials as $\conj{\zeta} = \operatorname{Re} \conj{\zeta} + \mathrm{i} \operatorname{Im} \conj{\zeta}$, so that the dequantization of the product \eref{eq:BakerCampbellHausdorff.ForWeylDequantization} reads
\begin{align}
\label{eq:BTDequantization.WeylDequantizationProduct}
            \varXi_{\hbar}\bigl(
                W_{\hbar}( \phi ) W_{\hbar}( \phi' )
            \bigr)
    &= \varXi_{\hbar}\bigl(
                W_{\hbar}( \phi )
            \bigr)
            \mathrm{e}^{-\conj{\zeta}}
            \varXi_{\hbar}\bigl(
                W_{\hbar}( \phi' )
            \bigr)
    .
\end{align}
Thus the dequantization of the product of sections becomes
\begin{subequations}
\begin{align}
\label{eq:CField.BTDequantizationSection.Product}
            ( A_{f} A_{f'} )( \hbar )
    &= \iint_{\mathcal{S}^*}
                \hat{f}( \phi, \hbar )
                \hat{f}'( \phi', \hbar )
                W_{\hbar}( \phi )
                W_{\hbar}( \phi' )
            \,\mathrm{d^2vol}^*
    , \allowdisplaybreaks[3]\\
\label{eq:CField.BTDequantizationSection.ProductDequantization}
            \varXi_{\hbar}\bigl( ( A_{f} A_{f'} )( \hbar ) \bigr)
    &= \iint_{\mathcal{S}^*}
                \hat{f}( \phi, \hbar )
                \hat{f}'( \phi', \hbar )
                \varXi_{\hbar}\bigl(
                    W_{\hbar}( \phi )
                \bigr)
                \mathrm{e}^{-\conj{\zeta}}
                \varXi_{\hbar}\bigl(
                    W_{\hbar}( \phi' )
                \bigr)
            \,\mathrm{d^2vol}^*
    .
\end{align}
\end{subequations}
The exponential $\mathrm{e}^{-\conj{\zeta}}$ is the Fourier transform of the derivatives that act on the $\mathrm{e}^{\mathrm{i} \phi}$ and $\mathrm{e}^{\mathrm{i} \phi'}$ functions of the Weyl generator dequantizations,
\begin{align}
\label{eq:CField.BTDequantizationSection.Intertwining}
            \mathrm{e}^{\mathrm{i} \phi}
            \mathrm{e}^{-\conj{\zeta}}
            \mathrm{e}^{\mathrm{i} \phi'}
    &= \mathrm{e}^{\mathrm{i} \phi}
            \left(
            \sum_{j = 0}^{\infty}
                \frac{1}{j!}
                \left(
                    \overset{\leftarrow}{\pderiv{}{\conj{z}^{\ib}}}
                    \delta^{\ib i}
                    \overset{\rightarrow}{\pderiv{}{z^{i}}}
                \right)^{j}
                \hbar^{j}
            \right)
            \mathrm{e}^{\mathrm{i} \phi'}
    .
\end{align}
Hence, the integration in \eref{eq:CField.BTDequantizationSection.ProductDequantization} separates into the integrals for the sections $A_{f}$ and $A_{g}$.
From the assumptions, we know that these sections are $\varXi$-expandable such that for any $k \in \mathbb{N}$:
\begin{align}
\label{eq:CField.BTDequantizationSection.Expansion}
            \Sigma^{k}_{\varXi}( A_{f}, \hbar )
    &= \sum_{j = 0}^{k} f_{j} \hbar^{j}
\end{align}
and similarly for $A_{f'}$.
We express the dequantization \eref{eq:CField.BTDequantizationSection.Dequantization} for both $A_{f}$ and $A_{f'}$ by the respective expansions \eref{eq:CField.BTDequantizationSection.Expansion} leading to
\begin{align}
        \Sigma^{\infty}_{\varXi}( A_{f}, \hbar )
        \exp\left(
            \hbar
            \overset{\leftarrow}{\pderiv{}{\conj{z}^{\ib}}}
            \delta^{\ib i}
            \overset{\rightarrow}{\pderiv{}{z^{i}}}
        \right)
        \Sigma^{\infty}_{\varXi}( A_{f'}, \hbar )
&= \Sigma^{\infty}_{\varXi}( A_{f}, \hbar )
        \star_{\varXi}
        \Sigma^{\infty}_{\varXi}( A_{f'}, \hbar )
    .
\end{align}

The dequantization star product is again Poisson compatible, self-adjoint and differential, which is analogously shown as in the previous proposition for the quantization star product.
\qed

The ``gauge transformation'' that relates the quantization star product $\star_{T}$ to the dequantization star product $\star_{\varXi}$, $\forall f_1, f_2 \in \mathrm{C}^\infty_\mathrm{S}( \mathcal{S}, \mathbb{C} )$:
\begin{align}
\label{eq:StarProduct.BerezinTransform}
			( \varXi_{\hbar} \after T_{\hbar} )
			\bigl( f_1 \star_{T} f_2 \bigr)
	&= \bigl(
				( \varXi_{\hbar} \after T_{\hbar} )( f_1 )
			\bigr)
		\star_{\varXi}
			\bigl(
				( \varXi_{\hbar} \after T_{\hbar} )( f_2 )
			\bigr)
	,
\end{align}
is the series expansion with the coefficients \eref{eq:BTSection.DequantizationExpandableCoeff}, determined by the expansion terms of the Berezin transform.

\section{Conclusion}
We considered the method of geometric quantization for a symplectic manifold with Riemannian metric \cite{2008MaMarinescu-Bergman}. When this method is applied to a symplectic vector space with an inner product, as is the case in QFT on causal sets, it naturally yields the Sorkin-Johnston state.
%In our approach to the Sorkin-Johnston state, we start with the given structure of a symplectic vector space with some inner product and use a natural geometric construction that yields the same state without imposing the axioms.
%This construction is based on a more general argument for symplectic manifolds that admit a Riemannian metric \cite{2008MaMarinescu-Bergman}.

For our case of a real, finite-dimensional vector space $\mathcal{S}$, we analysed the spectrum of the Bochner Laplacian of the quantization bundle to find the eigenspace of the lowest spectral value in \Sref{sec:VectorSpace.GeometricQuantization}.
This eigenspace is spanned by the holomorphic sections \eref{eq:QBundle.HolomSection} with respect to some complex structure $J$.
We choose this subspace of square-integrable sections as the physical Hilbert space $\mathcal{H}_{\hbar}$.
We showed that Toeplitz quantization gives a strict deformation quantization, which induces a star product \eref{eq:BTQuantization.StarProduct}.
The adjoint operation to Toeplitz quantization, referred to as dequantization, induces another star product \eref{eq:BTDequantization.StarProduct}. Dequantization maps quantum observables to classical observables, and by evaluation at $0$, this defines a state; we showed that this is precisely the  Sorkin-Johnston state.
%This geometric construction yields a complex structure, a compatible K\"ahler vector space, the algebra of quantum observables, and a state without imposing an axiomatic framework.

The above construction was done for a finite-dimensional symplectic vector space.
Such a finite dimensional system appears as the space of on-shell fields for the Klein-Gordon equation over a (subset) of a causal set (locally finite, partial ordered set) in causal set theory \cite{1987BombelliEtAl}.
We hope that our results will find applications in quantum field theory on causal sets, as well as in generalizations to symplectic manifolds, at least in those cases where the construction of quantum observables via geometric quantization is suitable.

Our construction also suggests a generalization to interacting theories, for which the phase space is no longer naturally described as a symplectic vector space. If a Riemannian metric is available on the phase space, then geometric quantization may be applied.
Berezin-Toeplitz dequantization and evaluation at 0, would then give a state that generalizes the Sorkin-Johnston state.

A further challenge is to extend this construction to fermionic systems and compare it with the construction of fermionic projector states \cite{1970Araki,2016FinsterMurroRoeken}.

\ack
C.M. would like to thank EPSRC (grant EP/N509802/1) for the funding of the PhD fellowship that made this research possible. E.H. would like to thank the Banff International Research Station, when this project was inspired. 
K.R. would like to thank the Perimeter Institute for hospitality and ongoing support. 
C.M. and K.R. also thank Fay Dowker, Rafael Sorkin and Sumati Surya for numerous inspiring discussions about causal sets. 

% \bmhead{Competing interests}
% On behalf of all authors, the corresponding author states that there is no conflict of interest.
% 
% \bmhead{Data availability}
% Data sharing not applicable to this article as no datasets were generated or analysed during the current study.
% 
% \bmhead{Authors' contributions}
% Not applicable
% 
% \bmhead{Ethics approval}
% Not applicable
% 
% \bmhead{Consent to participate}
% Not applicable
% 
% \bmhead{Consent for publication}
% Not applicable
% 
% \bmhead{Code availability}
% Not applicable

\appendix
\section{Normal and anti-normal ordering, and the Berezin transform}
\label{sec:AppendixNormalAntiNormalOrdering}
In this appendix, we want to demonstrate how the Berezin-Toeplitz quantization and dequantization relate to anti-normal and normal ordering, respectively.
Heuristically, we may extend the Toeplitz quantization map \eref{eq:BTQuantization} from continuous, unbounded functions to unbounded operators.
In particular for the coordinate functions (with any $i \in [ 1, N ]$), we may write
\begin{align}
\label{eq:BTQuantization.Coordinates}
			T_{\hbar}\bigl( \delta_{i \ib} z^{i} \bigr)
	&= \sqrt{\hbar} a^{+}_{\ib}
	&
			T_{\hbar}\bigl( \delta_{i \ib} \conj{z}^{\ib} \bigr)
	&= \sqrt{\hbar} a^{-}_{i}
	.
\end{align}
The dequantization map $\varXi$ may be extended in a similar way, giving
\begin{align}
\label{eq:BTDequantization.Coordinates}
			\varXi_{\hbar}\left( \sqrt{\hbar} a^{+}_{\ib} \right)
	&= \delta_{i \ib} z^{i}
	&
			\varXi_{\hbar}\left( \sqrt{\hbar} a^{-}_{i} \right)
	&= \delta_{i \ib} \conj{z}^{\ib}
	.
\end{align}
For a monomial, we have
\begin{align}
\label{eq:BTQuantization.Monomial}
			T_{\hbar}\Bigl(
				\left( \delta_{i \ib} z^{i} \right)^{m}
				\left( \delta_{j \jb} \conj{z}^{j} \right)^{n}
			\Bigr)
	&= \hbar^{\frac{m + n}{2}}
			\left( a^{-}_{j} \right)^{n}
			\left( a^{+}_{\ib} \right)^{m}
\end{align}
and
\begin{align}
\label{eq:BTDequantization.Monomial}
			\varXi_{\hbar}\Bigl(
				\hbar^{\frac{m + n}{2}}
				\left( a^{+}_{\ib} \right)^{m}
				\left( a^{-}_{j} \right)^{n}
			\Bigr)
	&= \left( \delta_{i \ib} z^{i} \right)^{m}
			\left( \delta_{j \jb} \conj{z}^{\jb} \right)^{n}
	,
\end{align}
both extending to any polynomials by linearity.
Notice the correspondence of quantization with anti-normal ordering and dequantization with normal ordering, respectively.

As a less heuristic example, consider a Schwartz function $f \in \mathrm{C}^\infty_\mathrm{S}( \mathcal{S}, \mathbb{C} )$ that is an $N$-fold product of Gaussian functions with variances $\beta_{i} > \hbar$ for all $i \in [ 1, N ]$, given by
\begin{subequations}
\label{eq:Gaussian.Function}
\begin{align}
			f( z )
	&:= \prod_{i = 1}^{N}
				\frac{1}{2 \pi \beta_{i}}
				\exp\left( - \frac{( x^{i} )^2 + ( y^{i} )^2}{\beta_{i}} \right)
	\\
	&= \prod_{i = 1}^{N}
				\frac{1}{2 \pi \beta_{i}}
				\exp\left( - \frac{| z^{i} |^2}{\beta_{i}} \right)
	.
\end{align}
\end{subequations}
It expands as a product of Taylor series
\begin{align}
\label{eq:Gaussian.Expansion}
			f( z )
	&= \prod_{i = 1}^{N}
				\frac{1}{2 \pi \beta_{i}}
				\sum_{k = 0}^{\infty}
					\frac{1}{k!}
					\left( - \frac{1}{\beta_{i}} \right)^{k}
					\left( z^{i} \right)^{k}
					\left( \conj{z}^{i} \right)^{k}
\end{align}
and its Toeplitz operator may be expanded similarly in terms of the unbounded ladder operators,
\begin{align}
\label{eq:Gaussian.BTQuantization}
			T_{\hbar}( f )
	&= \prod_{i = 1}^{N}
				\frac{1}{2 \pi \beta_{i}}
				\sum_{k = 0}^{\infty}
					\frac{1}{k!}
					\left( - \frac{\hbar}{\beta_{i}} \right)^{k}
					\left( a^{-}_{i} \right)^{k}
					\left( a^{+}_{i} \right)^{k}
	.
\end{align}
Note that the $k$-th power of the lowering operator appears to the left of the $k$-th power of the raising operator.
Thus the function $f$ is the anti-normal ordering corresponding to the Toeplitz operator $T_{\hbar}( f )$.

Now consider an observable $A$ with a similar expansion as \eref{eq:Gaussian.BTQuantization}, but with the opposite ordering of the ladder operators,
\begin{align}
\label{eq:Gaussian.Operator}
			A
	&:= \prod_{i = 1}^{N}
				\frac{1}{2 \pi \beta_{i}}
				\sum_{k = 0}^{\infty}
					\frac{1}{k!}
					\left( - \frac{\hbar}{\beta_{i}} \right)^{k}
					\left( a^{+}_{i} \right)^{k}
					\left( a^{-}_{i} \right)^{k}
	.
\end{align}
In contrast to \eref{eq:Gaussian.BTQuantization}, here the $k$-th power of the creation operator $( a^{+}_{i} )^{k}$ appear to the left of the $k$-th power of the annihilation operator $( a^{-}_{i} )^{k}$.
The $\varXi$-dequantization of the operator $A$ is
\begin{align}
\label{eq:Gaussian.BTDequantization}
			\varXi_{\hbar}( A )
	&= f
	.
\end{align}
Thus the operator $A$ corresponds to the function $f$ by normal-ordering.
This example shows the correspondence of Toeplitz quantization to anti-normal, and Toeplitz dequantization to normal ordered expressions.

Taking the dequantization of the Toeplitz operator $T_{\hbar}( f )$, we obtain the Berezin transform, here written with the convolution $\circledast$ as defined in \eref{eq:VectorSpace.Convolution},
\begin{subequations}
\label{eq:Gaussian.BTransform}
\begin{align}
			( \varXi_{\hbar} \after T_{\hbar} )( f )( z )
	&= \frac{1}{( 2 \pi \hbar )^{N}}
			\exp\left( -\frac{1}{\hbar} | z |^2 \right)
			\circledast f( z )
	\\\nonumber
	&=
			\frac{1}{( 2 \pi \hbar )^{N}}
			\int_{\mathcal{S}}
				\exp\left( -\frac{1}{\hbar} | \zeta |^2 \right)
 	\\
 	&\qquad\qquad\times
				\prod_{i = 1}^{N}
					\frac{1}{2 \pi \beta_{i}}
					\exp\left(
					- \frac{1}{\beta_{i}} \left| z^{i} - \zeta^{i} \right|^2
					\right)
			\,\mathrm{dvol}( \zeta )
	.
\end{align}
\end{subequations}
Because our Gaussian function $f$ is a product of independent Gaussian functions, the $2 N$-fold integration splits into $N$ double-integrals which are solved by completing the square in the exponents,
\begin{subequations}
\label{eq:Gaussian.BTransformSolution}
\begin{align}
	&= \prod_{i = 1}^{N}
				\frac{1}{( 2 \pi \beta_{i} ) ( 2 \pi \hbar )}
				\iint
					\exp\left(
					- \frac{1}{\hbar} \left| \zeta^{i} \right|^2
					- \frac{1}{\beta_{i}} \left| z^{i} - \zeta^{i} \right|^2
					\right)
					\mathrm{i} \,\mathrm{d}\zeta^{i} \mathrm{d}\conj{\zeta}^{i}
	\\\nonumber
	&= \prod_{i = 1}^{N}
			\Biggl(
				\frac{2 \pi \frac{\beta_{i} \hbar}{\beta_{i} + \hbar}}{( 2 \pi \beta_{i} ) ( 2 \pi \hbar )}
				\exp\left( - \frac{| z^{i} |^2}{\beta_{i} + \hbar} \right)
	\\
	&\qquad\times
				\underbrace{\iint
					\frac{1}{2 \pi \frac{\beta_{i} \hbar}{\beta_{i} + \hbar}}
					\exp\left(
					- \frac{\beta_{i} + \hbar}{\beta_{i} \hbar}
						\left| \zeta^{i} - \frac{\hbar}{\beta_{i} + \hbar} z^{i} \right|^2
					\right)
					\mathrm{i} \,\mathrm{d}\zeta^{i} \mathrm{d}\conj{\zeta}^{i}}_{= 1}
			\Biggr)
	\\
	&= \prod_{i = 1}^{N}
				\frac{1}{2 \pi ( \beta_{i} + \hbar )}
				\exp\left( - \frac{| z^{i} |^2}{\beta_{i} + \hbar} \right)
	.
\end{align}
\end{subequations}
The Berezin transform of the Gaussian function $f$ is again a Gaussian function with variances increased by $\hbar$.

\printbibliography

\end{document}